\documentclass[paper,a4paper,11pt,twoside]{JHEP3}
\usepackage{macros}
\usepackage{amssymb,amsmath}
\usepackage{epsfig}
\usepackage{psfrag}
\title{%
Non-perturbative tests of heavy quark effective theory
}

\author{%
\begin{flushleft}
\vspace{-0.5cm}
\vbox{
\epsfxsize=2.5 true cm
\epsfbox{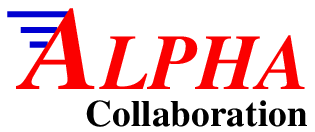}}
\end{flushleft}
}

\author{%
Jochen Heitger\\
Westf\"alische Wilhelms-Universit\"at M\"unster,
Institut f\"ur Theoretische Physik,\\
Wilhelm-Klemm-Strasse~9, D-48149 M\"unster, Germany\\
E-mail: \email{heitger@uni-muenster.de}
}
\author{%
Andreas J\"uttner\\
Humboldt Universit\"at Berlin, Institut f\"ur Physik,\\
Newtonstrasse~15, D-12489 Berlin, Germany\\
E-mail: \email{juettner@phys.soton.ac.uk}
}
\author{%
Rainer Sommer\thanks{%
On leave from: 
DESY Zeuthen, Platanenallee~6, D-15738 Zeuthen, Germany}\\
CERN, Theory Division, CH-1211 Geneva 23, Switzerland\\
E-mail: \email{rainer.sommer@desy.de}
}
\author{%
Jan Wennekers\\
Deutsches Elektronen-Synchrotron DESY, Theory Group,\\ 
Notkestrasse~85, D-22603 Hamburg, Germany\\
E-mail: \email{jan.wennekers@desy.de}
}

\preprint{%
MS-TP-04-11\\
HU-EP-04/33\\
CERN-PH-TH/2004-114\\
DESY 04-102\\
SFB/CPP-04-19\\
\today
}

\abstract{%
We consider QCD with one massless quark and one heavy quark in a 
finite volume of linear extent $L_0 \approx 0.2\,\fm$. 
In this situation, HQET represents an expansion in terms of 
$1/z=1/(m L_0)$, which we test by a non-perturbative computation
of quenched current matrix elements and energies, taking the 
continuum limit of lattice results.
These are seen to approach the corresponding renormalization group 
invariant matrix elements of the static effective theory as the 
quark mass becomes large. 
We are able to obtain estimates of the size of the 
$1/m$-corrections to the static theory,
which are also of practical relevance in our recent strategy to  
implement HQET non-perturbatively by matching to QCD in a
finite volume.
}

\keywords{%
HQET, Small-volume QCD, Matching, Size of $1/m$-corrections}

\begin{document}
\section{Introduction}
\label{Sec_intro}
It is expected that the dynamics of QCD simplifies in the limit of 
large masses of the charm and beauty quarks.
For external states with a single heavy quark, transition amplitudes 
are expected to be described by an effective quantum field theory, 
the Heavy Quark Effective Theory 
(HQET)~\cite{stat:eichhill1,hqet:polwis,hqet:georgi}. 
For details of the proper kinematics where this theory applies, and 
also for a guide to the original literature we refer the reader to
reviews \cite{HQET:neubert,reviews:HQETMannel}. 
To prepare for the following presentation, we only note that HQET applies
to matrix elements between hadronic states, where these hadrons are 
both at rest and do not represent high excitations.
HQET then provides an expansion of the QCD amplitudes in terms of $1/m$, 
the inverse of the heavy quark mass(es). 
The HQET Lagrangian of a heavy quark is given by\footnote{
We write here the velocity-zero part, since non-vanishing velocities
will not be relevant to our discussion.}
\bes
  \label{e_lag}
  \lag{HQET}(x) = %
  \heavyb(x)\left[\,D_{0}+ m 
                  - \frac{\omega_{\rm kin}}{2m}\,\vecD^2 
                  - \frac{\omega_{\rm spin}}{2m}\,\vecsigma\cdot \vecB 
            \,\right]\heavy(x) \,+\,\ldots \,,
\ees
where the ellipsis stands for higher-dimensional operators with 
coefficients of $\rmO(1/m^2)$. 
Following power counting arguments, this effective theory is 
renormalizable at any finite order in 
$1/m$ \cite{HQET:renormI,HQET:renormII}. 
A significant number of perturbative matching computations have 
been carried out (see \cite{reviews:HQETpert,ChetGrozin} and references 
therein), in order to express the parameters of HQET 
($m,\omega_{\rm spin},\ldots$) in terms of the QCD parameters. 
The very possibility of performing this matching reflects good evidence 
that HQET does represent an effective theory for QCD.

Nowadays, HQET is a standard phenomenological tool to describe
decays of heavy-light hadrons and their transitions in terms of a set 
of hadronic matrix elements, which are usually determined from 
experiments. 
Its phenomenological success is illustrated, for example, by
the determination of the Cabibbo-Kobayashi-Maskawa matrix 
element $V_{\rm cb}$: its value determined from inclusive $b \to c$
transitions agrees with the one extracted from exclusive 
ones \cite{PDGlast,reviews:Stone03} and HQET enters in both determinations.  
Similarly, HQET hadronic matrix elements, such as 
$\lambda_1\propto
\omega_{\rm kin}\langle{\rm B}|-\heavyb\vecD^2\heavy|{\rm B}\rangle$,
extracted from different experiments tend to 
agree \cite{reviews:Stone03,CKM:CERN,Babar:hqetpar}.
As a small caveat concerning these phenomenological tests, we note that 
some of them involve both the beauty and the charm quark, and one may 
suspect that the effective theory is not very accurate for the latter. 

Additional, independent tests of HQET are thus of both theoretical 
and phenomenological interest. 
In \cite{zastat:pap2} some Euclidean correlation functions, which are gauge 
invariant and infrared-finite, were studied in perturbation theory.  
There it was verified at one-loop order that their (large-distance) 
$m\to\infty$ asymptotics is described by the correlation functions of the 
properly renormalized effective theory.
This comparison was performed after separately taking the continuum limit 
of the lattice regularized effective theory and of QCD. 

In general, lattice gauge theory calculations allow a variation of the 
heavy quark mass and the performance of {\em non}-perturbative tests. 
However, if one is interested in the comparison of QCD and HQET in the 
{\em continuum limit}, one has to first respect the condition 
\bes
  m \ll {1 \over a}\,,
\ees
and then to do an extrapolation to zero lattice spacing, $a=0$. 
Given the present restrictions in the numerical simulations of lattice 
field theories, a direct comparison at large mass can only be done in 
a finite volume of linear size significantly smaller than $1\,\fm$. 

Before discussing this further, we note that in the charm quark mass 
region the continuum limit can also be taken in a large 
volume \cite{Bernard:2001fz,mcbar:RS02,fds:JR03}.
In particular, a recent study concentrated on the decay constant $\Fds$, 
where many previous estimates found evidence for large $1/m$-corrections. 
After taking the continuum limit and non-perturbatively
renormalizing the lowest-order HQET, \Ref{lat03:Juri} finds only a rather 
small difference between lowest-order HQET and the QCD results. 
Although this investigation was restricted to the quenched approximation, 
it provides some further evidence of the usefulness of HQET
-- maybe even for charm quarks. For related work, 
see \cite{fb_wupp,reviews:beauty,reviews:hartmut97,El-Khadra:1998hq,
Aoki:1998ji,Bernard:1998xi,heavylight:Bec98,Bowler:2000xw,AliKhan:2000eg,
Lellouch:2000tw,lat01:ryan,mb:roma2,fb:roma2c}
and references therein.

In this paper we study the large-mass behaviour of correlation functions 
in a finite volume of size $L \times L^3$, with Dirichlet boundary
conditions in time and periodic boundary conditions in space, i.e.~we work
in the framework of the QCD \txtSF \cite{SF:LNWW,SF:stefan1}.
Keeping all distances in the correlation functions of order $L$, 
the energy scale $1/L$ takes over the r\^ole usually played by small 
(residual) momenta, and at fixed $L$, HQET can be considered to be an 
expansion of QCD in terms of the dimensionless variable
\bes
  {1 \over z} \equiv {1 \over M L}\,,
\ees
where we take $M$ to be the renormalization group invariant (RGI) mass of 
the heavy quark (see~\eq{e_M}).\footnote{
We continue to use $m$ as a generic symbol for the quark mass when its 
precise definition is irrelevant, sometimes even not distinguishing the 
renormalized mass from the bare one. However, when precisely defined
functions of the quark mass are considered, only the properly defined 
$M$ is used. 
}
With the choice $L \approx 0.2\,\fm$, today's lattice techniques allow us 
to increase $z$ beyond $z=10$, while the continuum limit can still be
controlled well \cite{lat01:mbstat,HQET:pap2}. 
One is thus able to verify that the large-$z$ behaviour is described by 
the effective theory, which is the primary purpose of this paper.
Of course, the coefficients $a_i$ in expansions 
$\Phi=a_0+a_1/z+a_2/z^2+\ldots$ are functions of $\Lambda L$ 
(with $\Lambda$ the intrinsic QCD scale) such that 
$a_i(\Lambda L)\to c_i\times(\Lambda L)^i$ and 
$a_i(\Lambda L)/z^i\to c_i\times(\Lambda/M)^i$ as $\Lambda L\to\infty$.
Since we only work at one value of $L$, the dependence on $\Lambda L$ will 
be suppressed in the rest of this paper.

After a discussion of the QCD observables under investigation  
(\sect{Sec_def}), we give their HQET expansion (\sect{Sec_HQET})
and confront them with Monte Carlo results at finite values of $1/z$ 
(\sect{Sec_res}).
In our conclusions we also discuss the usefulness of our results
for a non-perturbative matching of HQET to QCD \cite{HQET:pap1}.

\section{Observables}
\label{Sec_def}
We introduce our observables starting from correlation functions defined 
in the continuum \txtSF (SF) \cite{SF:LNWW,SF:stefan1}. 
For the reader who is unfamiliar with this setting, we give a 
representation of these correlation functions in terms of operator matrix 
elements below.

We take a $T\times L^3$ geometry with $T/L=\rmO(1)$ fixed. 
The boundary conditions are periodic in space, where for the 
fermion fields, and only for those, a phase is introduced:
\bes
  \psi(x+\hat{k}L) = \rme^{i \theta} \psi(x)\,,\quad 
  \psibar(x+\hat{k}L) = \psibar(x)\rme^{-i \theta} \,,
  \quad k=1,2,3\,.
\ees
In the numerical investigation of \sect{Sec_res}, we will set $T=L$
and $\theta=0.5$.
Dirichlet conditions are imposed at the $x_0=0$ and $x_0=T$ boundaries. 
Their form as well as the action are by now 
standard \cite{SF:LNWW,SF:stefan1,zastat:pap1} and we do not repeat them
here. 
Multiplicatively renormalizable, gauge-invariant correlation functions 
can be formed from local composite fields in the interior of the manifold 
and from boundary quark fields. Boundary fields, located at the $x_0=0$ 
surface are denoted by $\zetabar,\zeta$ and create fermions and 
anti-fermions. Their partners at $x_0=T$ are $\zetabarprime,\zetaprime$. 
We shall need flavour labels,
``$\rm l$'' denoting a massless flavour,
``$\rm b$'' a heavy but finite-mass flavour and
``$\rm h$'' the corresponding field in the effective theory. 
The heavy-light axial vector and vector currents then read
\bes
  A_\mu(x)
   = 
  \lightb(x)\gamma_\mu\gamma_5\psi_{\rm b}(x)
  \,, \quad
  V_\mu(x)
   = 
  \lightb(x)\gamma_\mu\psi_{\rm b}(x)
  \,.
\ees
\subsection{Correlation functions \label{s_corr}}
In our tests we shall consider the correlation functions
\bes
  \fa(x_0) &=& -{1 \over 2}\int{\rmd^3\vecy\, \rmd^3\vecz}\,
  \left\langle
  A_0(x)\,\zetabar_{\rm b}(\vecy)\gamma_5\zeta_{\rm l}(\vecz)
  \right\rangle  \,, \label{e_fa} \\
  \kv(x_0) &=& -{1 \over 6}\sum_{k}\int{\rmd^3\vecy\, \rmd^3\vecz}\,
  \left\langle
  V_k(x)\,\zetabar_{\rm b}(\vecy)\gamma_k\zeta_{\rm l}(\vecz)
  \right\rangle  \,, \label{e_kv}
\ees
as well as the boundary-to-boundary correlations
\bes
  \fone &=& -{1 \over 2L^6}\int{\rmd^3\vecy\, \rmd^3\vecz\,
                             \rmd^3\vecy'\,\rmd^3\vecz'}\,
  \left\langle
  \zetabarprime_{\rm l}(\vecy')\gamma_5\zetaprime_{\rm b}(\vecz')\;
  \zetabar_{\rm b}(\vecy)\gamma_5\zeta_{\rm l}(\vecz)
  \right\rangle  \,, \label{e_fone} \\
  \kone &=& -{1 \over 6L^6}\sum_{k}\int{\rmd^3\vecy\, \rmd^3\vecz\,
                             \rmd^3\vecy'\,\rmd^3\vecz'}\,
  \left\langle
  \zetabarprime_{\rm l}(\vecy')\gamma_k\zetaprime_{\rm b}(\vecz')\;
  \zetabar_{\rm b}(\vecy)\gamma_k\zeta_{\rm l}(\vecz)
  \right\rangle  \,. \label{e_kone}
\ees
They are illustrated in \fig{fig:f_correl}.
%
\FIGURE{
\hspace{1.0cm}
\begin{minipage}{.3\linewidth}
\vspace{3mm}
\psfrag{1}[c][c][1][0]{$\overline{\zeta}_{\rm b}$}
\psfrag{2}[c][c][1][0]{$\zeta_{\rm l}$}
\psfrag{3}[c][r][1][0]{${A}_0(x_0)$}
\psfrag{4}[r][t][1][0]{$x_0=0$}
\psfrag{5}[r][r][1][0]{$x_0=T$}
\psfrag{7}[c][c][1][0]{\large$f_{\rm A}$}
\psfrag{8}[c][c][1][0]{$\gamma_5$}
\epsfig{scale=.45,file=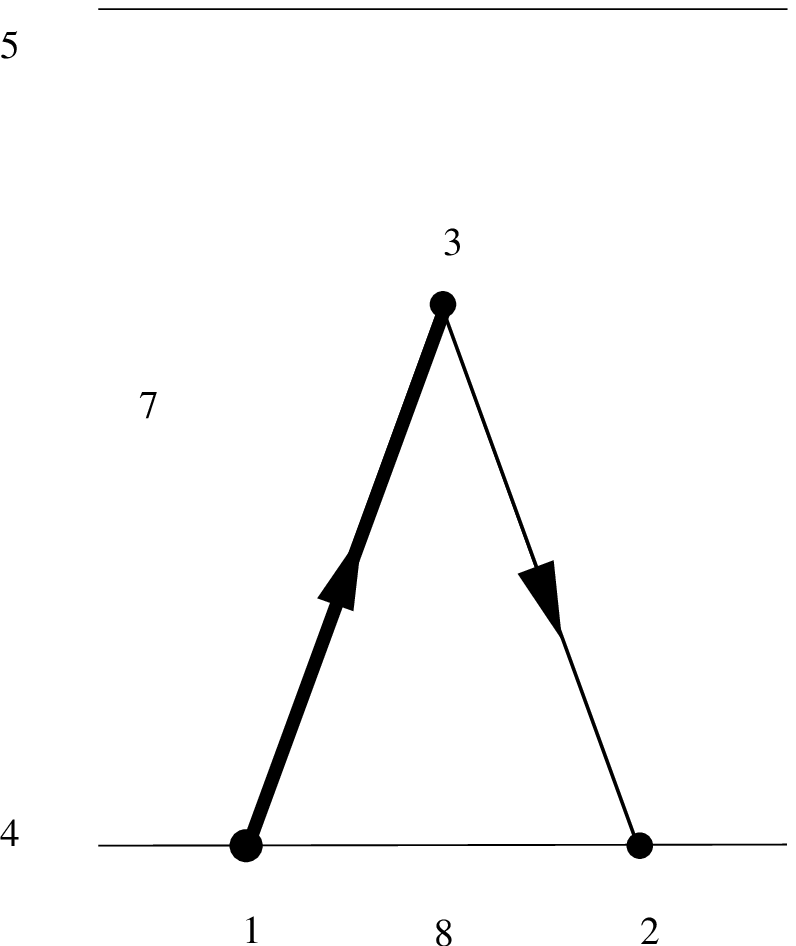}
\end{minipage}
\begin{minipage}{.3\linewidth}
\vspace{3mm}
\psfrag{1}[c][c][1][0]{$\overline{\zeta}_{\rm b}$}
\psfrag{2}[c][c][1][0]{$\zeta_{\rm l}$}
\psfrag{3}[c][r][1][0]{${V}_k(x_0)$}
\psfrag{4}[r][t][1][0]{}
\psfrag{5}[r][r][1][0]{}
\psfrag{7}[c][c][1][0]{\large$k_{\rm V}$}
\psfrag{8}[c][c][1][0]{$\gamma_k$}
\epsfig{scale=.45,file=plots/f_A.eps}
\end{minipage}
\begin{minipage}{.3\linewidth}
\vspace{-1.15mm}
\psfrag{1}[c][c][1][0]{$\overline{\zeta}_{\rm b}$}
\psfrag{2}[c][c][1][0]{$\zeta_{\rm l}$}
\psfrag{3}[c][c][1][0]{$\overline{\zeta}^\prime_{\rm l}$}
\psfrag{6}[c][c][1][0]{$\zeta^\prime_{\rm b}$}
\psfrag{4}[r][t][1][0]{}
\psfrag{5}[r][r][1][0]{}
\psfrag{7}[c][c][1][0]{\large$f_1$}
\psfrag{8}[c][c][1][0]{$\gamma_5$}
\psfrag{9}[c][c][1][0]{$\gamma_5$}
\epsfig{scale=.45,file=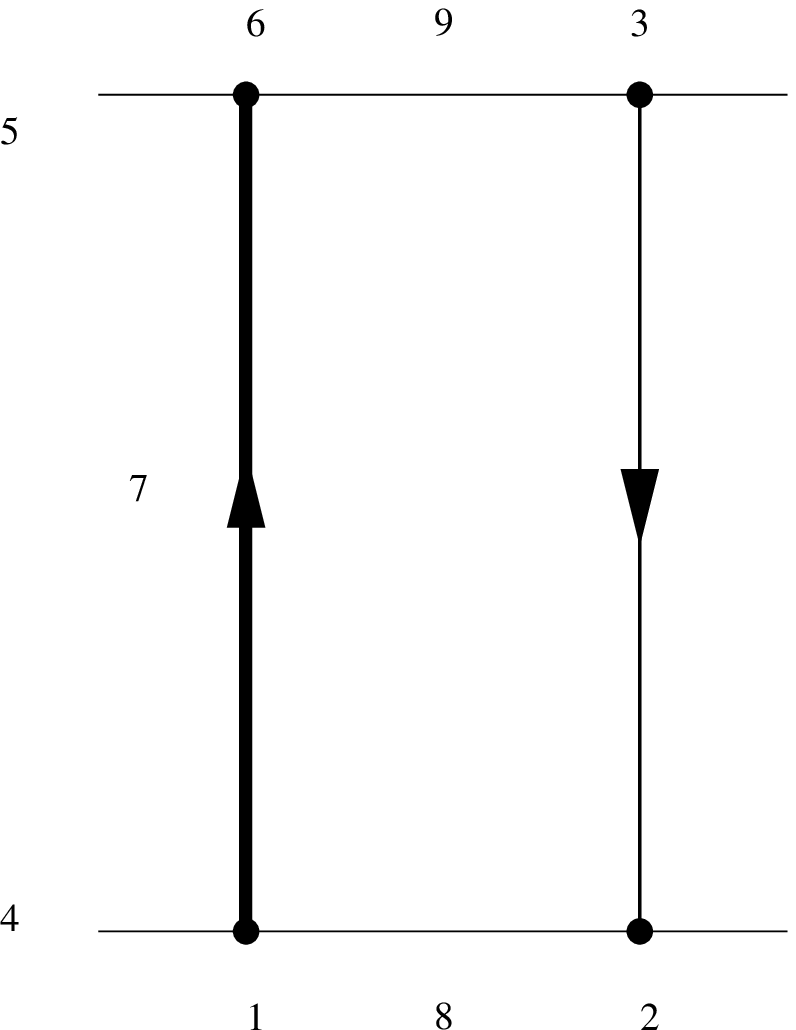}
\end{minipage}
\vspace{0.125cm}
\caption{
The Schr\"odinger functional correlation functions $f_{\rm A}$, 
$k_{\rm V}$ and $f_1$.
For $\kone$, in the rightmost diagram $\gamma_5$ is replaced by 
$\gamma_k$.
}\label{fig:f_correl}
}
%

Let us discuss the function $\fa(x_0)$ in some detail.  
It describes the creation of a (finite-volume) $\vecp=0$ heavy-light \ps 
meson state, $|\varphi_{\rm B}(L)\rangle$, through quark and antiquark
boundary fields which are separately projected onto 
momentum $\vecp=0$. 
``After'' propagating for a Euclidean time interval $x_0$, the axial 
current operator $\opA$ initiates a transition to a state, 
$|\Omega'(L)\rangle$, with the quantum numbers of the vacuum. 
The correlation function $\fa(x_0)$, taken in the middle of the manifold,
can thus be written in terms of Hilbert space matrix elements,
\bes
  \fa(T/2) = {\cal Z}^{-1} \langle \Omega(L)| \opA |B(L)\rangle
  \equiv {\cal Z}^{-1} \langle \Omega(L)| \Omega'(L)\rangle\,,\quad
  |B(L)\rangle = \rme^{-T\ham /2 } |\varphi_{\rm B}(L)\rangle \,,
\ees
where $\ham$ is the QCD Hamiltonian and 
\bes
  {\cal Z} =  \langle \Omega(L)| \Omega(L)\rangle\,,  \quad 
  |\Omega(L)\rangle =\rme^{-T\ham/2 } |\varphi_{0}(L)\rangle \,.
\ees
Here, $|\varphi_{0}(L)\rangle$ denotes the \txtSF intrinsic boundary
state. It has the quantum numbers of the vacuum. 
Because the definition of our correlation functions contains integrations 
over all spatial coordinates, all states appearing in our analysis are 
eigenstates of spatial momentum with eigenvalue zero.
The operator $\rme^{-T\ham/2 }$ suppresses high-energy states. 
When expanded in terms of eigenstates of the Hamiltonian,
$| \Omega(L)\rangle$ and $|B(L)\rangle$ are thus dominated by 
contributions with energies of at most $\Delta E = \rmO( 1/L)$ above the 
ground state energy in the respective channel (recall that we take 
$T=\rmO(L)$). 
This explains why, at large time separation $x_0\geq \rmO(1/m)$, 
HQET is expected to describe the large-mass behaviour of the correlation 
function, also in the somewhat unfamiliar framework of the 
Schr\"odinger functional.\footnote{ 
More generally, HQET will apply to correlation functions at large 
Euclidean separations. 
}

Equations similar to the above hold for $\kv$; one only needs to replace 
\ps states by vector ones. 
Finally, the boundary-to-boundary correlator may be represented as
\bea
 \fone = {\cal Z}^{-1} \langle B(L)| B(L)\rangle \,.
\eea
Since the boundary quark fields $\zeta,\zetabar,\ldots$ are 
multiplicatively renormalizable \cite{SF:stefan2}, this holds also for 
the states $|\varphi_{0}(L)\rangle$ and $|\varphi_{\rm B}(L)\rangle$.

It now follows that the ratios
\bea
\Yr(L,M) \equiv \frac{\fa(T/2)}{\sqrt{\fone}} \,,\quad 
\Yv(L,M) \equiv - \frac{\kv(T/2)}{\sqrt{\kone}} \,,\quad 
\Rr(L,M) \equiv - \frac{\fa(T/2)}{\kv(T/2)} \label{e_ratios}
\eea
are finite quantities when we adopt the convention that $A_\mu, V_\mu$ 
denote the renormalized currents. 
As is immediately clear from the foregoing discussion,
\bea
  \Yr(L,M) = { \langle \Omega(L)| \opA |B(L)\rangle 
		\over 
		||\;| \Omega(L)\rangle\;|| \;\;
		||\;| B(L)\rangle\;|| }
\eea
(\,or $\Yv(L,M)$\,) becomes proportional to the \ps (or vector) 
heavy-light decay constant as $L\to\infty$. 
We shall study the large-$M$ behaviour of these quantities at fixed $L$ 
in the following sections.

For the same purpose we define effective energies
\bea
\gamps(L,M)
& \equiv &
\left. 
-\frac{\rmd }{\rmd x_0} 
\ln\left[\,\fa(x_0)\,\right]\,\right|_{\,x_0=T/2}\,=\, 
-\frac{\fa'(T/2)}{\fa(T/2)}\,,
\label{gamps}\\
\gamv(L,M)
& \equiv &
\left.
-\frac{\rmd }{\rmd x_0}
\ln\left[\,\kv(x_0)\,\right]\,\right|_{\,x_0=T/2}\,=\, 
- \frac{\kv'(T/2)}{\kv(T/2)}\,.
\label{gamv}
\eea
These may be written in terms of matrix elements of the Hamiltonian. 
E.g., we have
\bea
\gamps(L,M) = { \langle B'(L)|\,\ham\,| B(L)\rangle
            \over
              \langle B'(L)| B(L)\rangle}
           - { \langle \Omega(L)|\,\ham\,| \Omega'(L)\rangle
            \over
              \langle \Omega(L)| \Omega'(L)\rangle}\,,
\eea
with $|B'(L)\rangle = \opA^\dagger | \Omega(L)\rangle$ and
$\langle B'(L)| B(L)\rangle =  \langle \Omega(L)| \Omega'(L)\rangle$.
Expanding in terms of energy eigenfunctions, one sees immediately that 
$\gamps(L,M) = \sum_i \beta_i E_B^{(i)} - \sum_i \alpha_i E_\Omega^{(i)}$,
where $E_B^{(i)}$ are the (finite-volume) energy levels in the heavy-light 
\ps meson sector and $E_\Omega^{(i)}$ those with vacuum quantum numbers. 
The coefficients $\beta_i,\alpha_i$ have a strong dependence on $i$, which 
labels the energy excitations; states with $E^{(i)}-E^{(0)}\gg 1/L$ are 
suppressed exponentially in the sum. 
For $z=ML \gg 1$, the effective energy $\gamps(L,M)$ is hence expected 
to be given by HQET. 

To summarize, $\Yr$, $\Yv$ and $\Rr$ are (ratios of) matrix elements
between low-energy heavy-light and vacuum-like states.
HQET at order $(1/m)^n$ should describe them up to corrections of the
order of $1/(ML)^{n+1}$. 
The energies $\Gamma$, eqs.~(\ref{gamps}) and (\ref{gamv}), share the same 
property. 
It is then possible to test HQET by studying the large-$z$ asymptotics
of these observables!

\section{Large-$z$ asymptotics: effective theory predictions}
\label{Sec_HQET}
We now turn our attention to the effective theory predictions 
for the observables introduced above. 
\subsection{Current matrix elements}
{\em At the classical level} it is expected that they can be described 
by a power series in $1/z$ with $z=ML$. 
This has been checked explicitly in \cite{zastat:pap2}: the expansion
in $1/z$ is asymptotic, non-analytic terms are of the type
$\,\rme^{\,-z}$ and are thus very small for, say, $z>4$. 
The leading term in the expansion for each of the correlation functions in 
\eqs{e_fa}--(\ref{e_kone}) is given by exactly the same expressions, 
evaluated with the simple replacement $\psi_{\rm b} \to \heavy$ etc.~and 
by dropping all terms of $\rmO(1/m)$ in the action associated 
with \eq{e_lag}. 
This corresponds to the static limit, where the heavy quark does not 
propagate in space.
We denote the corresponding observables (in the effective theory) with a 
superscript ``stat'', e.g. 
\bes \nonumber
  \fa(x_0) \to \fastat(x_0) \,,
\ees
and also introduce
\bes \label{e_X} 
  X(L) \equiv \frac{\fastat(T/2)}{\sqrt{\fonestat}} 
\ees
that is easily seen to be mass independent from the form of the static 
propagator (both in the formal continuum theory and in lattice 
regularization at finite lattice spacing \cite{zastat:pap1}). 
An example for the correspondence of the effective theory and QCD at the 
classical level is
\bes
  X(L)  
  = \lim_{z\to\infty} \Yr(L,M) = \lim_{z\to\infty} \Yv(L,M)\,.
\ees

{\em In the quantum theory} there are logarithmic modifications of such 
relations.
Of course, they have their origin in the scale dependent renormalization 
of the effective theory. 
For example, due to the renormalization of the axial current in the 
effective theory, the renormalized ratio\footnote{
One should {\em not} identify $\mu=1/L$ here, as it was done in the 
computation of the scale dependence of the static axial current 
in \cite{zastat:pap3}.
Note also that in the ratio $X$ other renormalization factors cancel 
in the effective theory, just as they do in QCD.
}
\bes
  \Xren(L,\mu) = \zastat(\mu)  \Xbare(L)\,
\ees
depends logarithmically on the chosen renormalization scale $\mu$.
It further depends on the chosen renormalization scheme, but the 
renormalization group invariant 
\be \label{e_XRGI}
  \XRGI(L) = 
  \lim_{\mu\to\infty}\,\left\{
  \left[\,2b_0\gbar^2(\mu)\,\right]^{-\gamma_0/(2b_0)}\,\Xren(L,\mu)\right\} 
  = \ZRGI \Xbare(L)\,, 
\ee
\be
  b_0=\frac{11}{(4\pi)^2}\,,\quad
  \gamma_0=-\frac{1}{(4\pi^2)} \qquad (\nf=0)\,,
\ee
does not. 
The renormalization constant $\ZRGI$ is computable in lattice 
QCD \cite{zastat:pap3}. 
Above, $b_0$ and $\gamma_0$ are given for the case of a vanishing number 
of flavours as is appropriate for the quenched approximation, which we
will employ in the following section.

The large-$z$ behaviour of the QCD observables is then given by the 
corresponding renormalization group invariants of the effective theory, 
together with logarithmically mass dependent functions that will be called
$C$ below. 
As arguments of these functions we choose the renormalization group 
invariant mass $M$ (in units of the $\Lambda$ parameter), since this can 
be fixed in the lattice computations without perturbative uncertainties: 
the relation between the bare quark masses in the lattice regularization,
which we use \cite{impr:pap3}, and $M$ has been non-perturbatively computed 
in \cite{msbar:pap1,impr:babp,HQET:pap2}.
The scheme independent $M$ describes the limiting behaviour of any running 
mass $\mbar(\mu)$ for large $\mu$ via
\bea \label{e_M}
  \lim_{\mu\to\infty}\,\left\{
  \left[\,2b_0\gbar^2(\mu)\,\right]^{-d_0/(2b_0)}\,\mbar(\mu)\right\}=M\,,
  \quad d_0=\frac{8}{(4\pi)^2} \qquad (\nf=0)\,.
\eea

The following predictions are then obtained 
(see also section~5.1 of \Ref{zastat:pap3}):
\bea
\Yr(L,M) & \simas{M\to\infty}\,\, & 
\Cps\left(M/\lMSbar\right)\,\XRGI(L)\,\Big(1 + \rmO(1/z)\Big)\,,
\quad z=ML\,, 
\label{yR2stat}\\
\Yv(L,M) & \simas{M\to\infty}\,\, & 
\Cv\left(M/\lMSbar\right)\,\XRGI(L)\,\Big(1 + \rmO(1/z)\Big)\,,
\label{yRv2stat}\\
\Rr(L,M) & \simas{M\to\infty}\,\, &
\Cpsv\left(M/\lMSbar\right)\,\Big(1 + \rmO(1/z) \Big)\,.
\label{rR2stat}
\eea
Here, the function $\Cps(M/\lMSbar)$ has the asymptotics
\bea
  \Cps\left(M/\lMSbar\right) \simas{M\to\infty}\,\, 
  \left(\ln\frac{M}\lMSbar \right)^{-\gamma_0/(2b_0)} 
  \left\{ 1 + \rmO\left(
  \frac{\ln\left[\ln\left(M/\lMSbar\right)\right]}
  {\ln\left(M/\lMSbar\right)}
  \right)\right\}\,;
\eea
to within this accuracy, $\Cv$ shares this asymptotic behaviour. 
In practice, the functions $C_{\rm X}(M/\lMSbar)$, 
${\rm X}={\rm PS,V,PS/V}$, are obtained by solving the perturbative 
renormalization group equations along the lines of section~5.1 of 
\Ref{zastat:pap3}, where more details can be found.
In particular we always use the four-loop perturbative approximation
of the $\beta$-function \cite{vanRitbergen:1997va} and the three-loop 
approximation to the anomalous dimension $\gamma$ of the currents, which 
has recently been computed \cite{ChetGrozin}. 
Taking the $n$-loop approximation to $\gamma$, the resulting relative 
error in the functions $C_{\rm X}$ is of order $\alpha^{n}$ with $\alpha$ 
evaluated at a scale of the order of the heavy quark mass. 
Explicit expressions for the functions $C_{\rm X}$ are given 
in \App{App_Cx}.
%
\FIGURE{
\epsfig{file=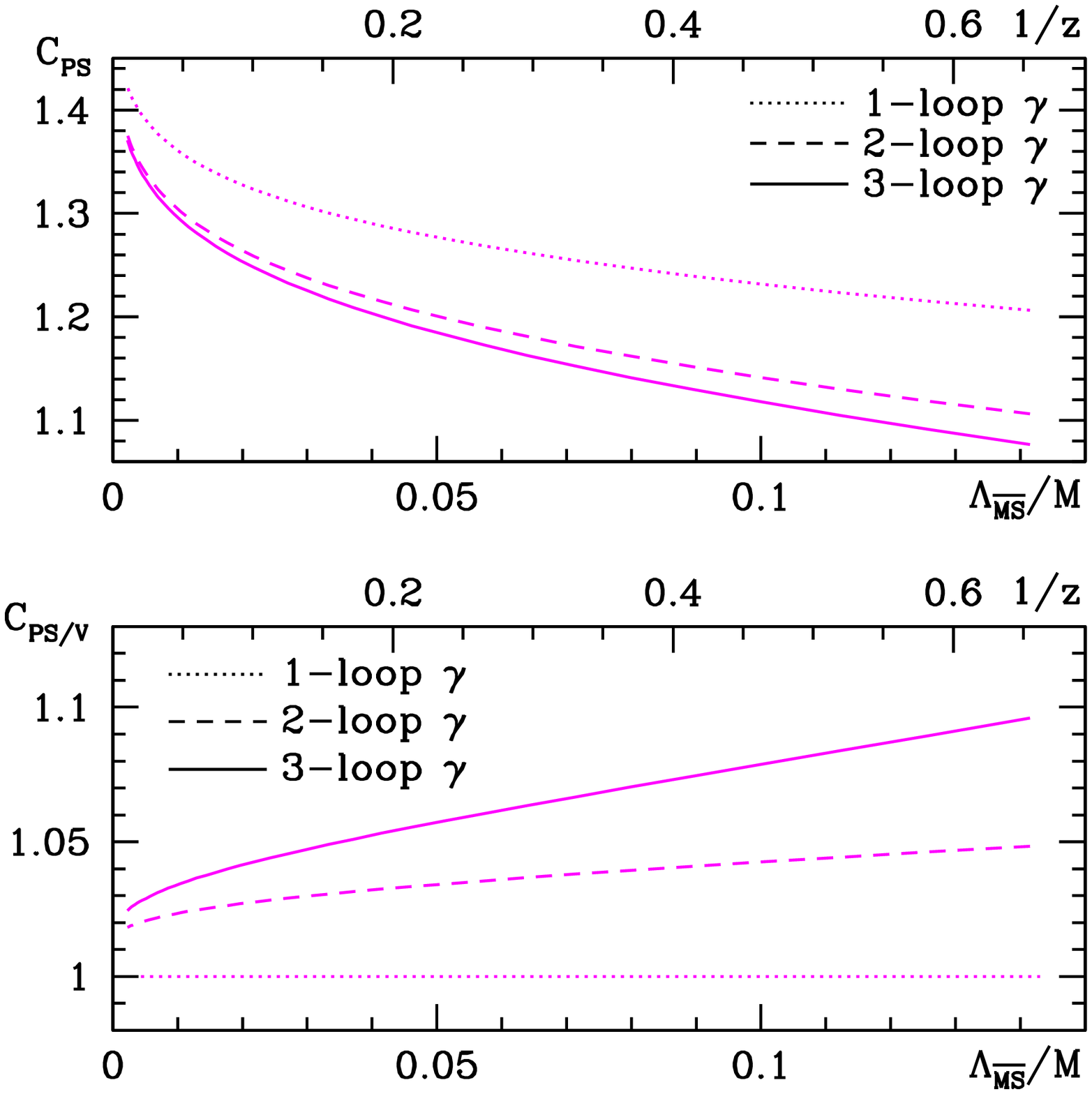,width=9cm} 
\vspace{1.0cm}
\caption{The functions $\Cps$ and $\Cpsv$ evaluated in different 
approximations of perturbation theory. 
In relating $\lMSbar/M$ to $z=ML$, we used \cite{msbar:pap1}
$\lMSbar\,r_0=0.602$ and $L/r_0=\Lmax/(2r_0)=0.359$.
(Setting $r_0=0.5\,\fm$, this actually corresponds to $L\approx0.2\,\fm$.)
}\label{f_Cpert}
}
%

In order to study whether their $\alpha^{n}$-error is a limiting factor,
we first note that changing the order from three-loop to four-loop in
the $\beta$-function makes only a tiny difference. 
Next we plot $C_{\rm X}$, ${\rm X}={\rm PS,PS/V}$, using $\gamma$ at 
$n$ loops in \fig{f_Cpert}.
We observe a reasonable behaviour of the asymptotic perturbative series 
and take half of the difference between the two-loop and three-loop 
approximations for $\gamma$ as our uncertainty. 
This uncertainty will be almost negligible with respect to our statistical 
errors. Note that without the three-loop computations of \cite{ChetGrozin} 
such a statement would not have been possible. 

For completeness we remind the reader that \eqs{yR2stat}--(\ref{rR2stat}) 
are afflicted by the usual problem of identifying power corrections. 
Asymptotically, at large $M$, the higher-order logarithmic (perturbative) 
corrections in $C_{\rm X}$ dominate over the $1/z^n$ power corrections. 
However, as just discussed, in the interesting range of $z$, the 
perturbative corrections are under reasonable control and it makes sense to 
investigate the power corrections if they are significantly larger. 
Note that this problem is not present if the effective theory is 
renormalized non-perturbatively \cite{HQET:pap1}.
\subsection{Energies \label{s_ener}}
Next we concentrate on the energies $\Gamma$. Clearly, because of the 
term $m \heavyb(x)\heavy(x)$ in the Lagrangian (\ref{e_lag}), 
they grow (roughly linearly) with the mass. 
For the case of hadron masses, mass formulae have been written 
down \cite{HQET:massform2}; for instance, the one for the B-meson mass 
reads:
\bes \label{e_massformula}
 \mB = m + \bar{\Lambda} + \frac1{2m}\,(\lambda_1 + d_{\rm B}\lambda_2 ) 
       \,+\, \rmO\left(1/m^2\right) \,,\quad
 d_{\rm B} = 3 
\ees
(and the same formula holds for the ${\rm B}^\ast$-meson except that 
$d_{\rm B} \to d_{{\rm B}^\ast} = -1$).
The matrix element
$\lambda_1=\half\,
\omega_{\rm kin}\langle {\rm B}| -\heavyb\vecD^2\heavy|{\rm B}\rangle$ 
was mentioned already in the introduction, and
$\lambda_2=\frac{1}{2 d_{\rm B}}\,\omega_{\rm spin}\langle {\rm B}|
-\heavyb\vecsigma\cdot\vecB\,\heavy|{\rm B}\rangle$. 
Some cautioning remark concerning the above formula is in order.  
It suggests that the binding energy $\bar{\Lambda}=\mB - m + \rmO(1/m)$ 
may be obtained as a prediction of HQET. 
However, there is no unique non-perturbative definition of the mass $m$, 
which should be subtracted. As a consequence, $\bar{\Lambda}$
has an ambiguity of order $\Lambda_{\rm QCD}$, which may then also
propagate into a significant ambiguity in $\lambda_1$ extracted 
from \eq{e_massformula}.

We nevertheless start our discussion of the effective energies 
$\Gamma$ from a trivial generalization of \eq{e_massformula}:
\bea \label{e_effenergy_exp}
\gamav(L,M) 
& \equiv &
\frac14\Big[\,\gamps(L,M)+3\,\gamv(L,M)\,\Big] \nonumber\\ 
&    =   &
m + \bar{\Lambda}(L) + \frac1{2m}\,\lambda_1(L) 
\,+\, \rmO\left(1/m^2\right) \,,
\eea 
where $\lambda_1(L)$ again summarizes the effect of the 
$\heavyb\vecD^2\heavy$-perturbation to the static action.
We have cancelled a $\lambda_2$-term by considering the spin-averaged 
combination of the energy. 
While one ought to be careful with the interpretation of sub-leading 
terms in $1/M$, the large-mass behaviour of $\gamav(L,M)$ is given by 
\be
 L\gamav(L,M) \,\simas{M\to\infty}\,\,\, 
 \Cmass\left(M/\lMSbar\right)\times z \,+\, \rmO\left((1/z)^0\right)\,, 
 \label{e_gamasympt}
\ee 
with (\,$\mbar_\msbar(\mbar_*)=\mbar_*$\,)
\bea \label{e_cmass}
\Cmass\left(M/\lMSbar\right) = {m_Q \over M} = 
       {\mbar_* \over  M}\, {m_Q \over \mbar_*}
\ees
and $m_Q$ being the pole mass. 
Here the first factor on the right-hand side is known very precisely in 
perturbation theory (up to four 
loops \cite{Chetyrkin:1997dh,Vermaseren:1997fq}), but it is well known 
that the perturbative series for the second factor is not very well 
behaved and even the three-loop term \cite{polemass:3loop} is still 
rather significant. 
We will discuss this uncertainty in $\Cmass$ together with the numerical
results in the following section.\footnote{
We note that \eqs{e_gamasympt} and (\ref{e_cmass}) are the only ones
where $m_Q$ enters the coefficient functions relating the RGI matrix 
elements of HQET to the QCD observables.
In all other cases, $m_Q$ has been eliminated and only $M$ appears.
Thus, there is no particular reason to expect that any of the perturbative 
expressions (for the various anomalous dimension functions) is badly 
behaved.
}

Let us now consider the sub-leading terms in \eq{e_effenergy_exp}.
Taking energy differences, $m$ drops out and 
e.g.~$\bar{\Lambda}(L)-\bar{\Lambda}(L')$ can be computed unambiguously 
from the static effective theory. 
It does not depend on any convention adopted for $m$ 
in \eq{e_effenergy_exp}. 
In our numerical computations, however, we investigated only one value
of $L$. As an example of another observable unaffected by the ambiguity 
in $m$, we therefore look at the combination
\bea
 \gamdif(L,M)= 
 \frac{L}{4}\left[\,
 \frac{\fa'(T/4)}{\fa(T/4)} - \frac{\fa'(T/2)}{\fa(T/2)} 
 +3\,\frac{\kv'(T/4)}{\kv(T/4)} - 3\,\frac{\kv'(T/2)}{\kv(T/2)}
  \,\right]\,.
\label{e_Xi}
\eea
The HQET prediction for this quantity is 
\bea\label{e_gamdif}
  \gamdif(L,M) = \gamdifstat(L) + \frac1{2z}\,\gamdifone(L) 
  \,+\, \rmO\left(1/z^2\right)\,,
\eea
with 
\bea\label{e_gamdifstat}
 \gamdifstat(L) = 
 L\left[\,
 \frac{(\fastat)'(T/4)}{\fastat(T/4)}-\frac{(\fastat)'(T/2)}{\fastat(T/2)}
 \,\right]\,,
\eea
defined in the static effective theory. 
Since it is an energy, $\gamdif$ does not require any renormalization.
The first-order correction in $1/z$ is given entirely by matrix elements 
of $\heavyb(x)\vecD^2\heavy(x)$. 
Its coefficient is fixed using the reparametrization invariance of the 
effective theory, first discussed in \Ref{HQET:reparaI}. 
To see this, one considers the matrix elements of 
$\heavyb(x)\vecD^2\heavy(x)$ to be computed in dimensional regularization. 
Then reparametrization invariance is valid, and the coefficient of the 
operator renormalized by minimal subtraction is the inverse $\msbar$ quark 
mass \cite{HQET:reparaII,HQET:reparaIII}, whose renormalization scale 
dependence cancels against the one of the matrix element. 
From \eq{e_M} it hence follows that the prefactor of the renormalization
group invariant matrix element of $\heavyb(x)\vecD^2\heavy(x)$ is $1/(2M)$, 
as used in \eq{e_gamdif}. 
In that equation, $\gamdifone(L)$ is the matrix element made dimensionless
by a factor $L$. In the following, the exact expression for 
$\gamdifone(L)$ will be irrelevant. It rather suffices to know that it 
does not depend on the mass.

We point out that, in writing down the quantum mechanical representation 
of $\gamdif$, states play a r\^ole for which only the operator
$\rme^{-T\ham/4}$ is effective to suppress higher energy contributions, 
while in our other observables $\rme^{-T\ham/2}$ suppresses high energies. 
For the quantity $\gamdif$, HQET is thus expected to be accurate only at 
larger values of $z$. 
This variable should roughly be a factor of 2 larger than for the other 
observables. 

Finally, in the difference
\be
\delgam(L,M) \,\equiv\, \gamps(L,M)-\gamv(L,M)
\label{delgam}
\ee
the lowest-order term that contributes in the effective theory is 
$\heavyb(x)\vecsigma\cdot\vecB\,\heavy(x)$. 
With $\Cspin$ constructed from the anomalous dimension of this operator 
in the effective theory, 
$\gamma^{\rm spin}$ \cite{HQET:sigmabI,HQET:sigmabII}, we thus have
\be
  L\delgam(L,M) \,\simas{M\to\infty}\,\,\, 
  \Cspin\left(M/\lMSbar\right)\,\frac{\XRGIspin(L)}{z}\,
  \Big(1+\rmO(1/z)\Big)\,, 
  \label{e_XRGIspin}
\ee
where the leading asymptotics of the function $\Cspin(M/\lMSbar)$ is of 
the form
\be
  \Cspin\left(M/\lMSbar\right) \simas{M\to\infty}\,\,
  \left(\ln\frac{M}\lMSbar \right)^{-\gamma_0^{\rm spin}/(2b_0)} 
  \left\{ 1 + \rmO\left(
  \frac{\ln\left[\ln\left(M/\lMSbar\right)\right]}
  {\ln\left(M/\lMSbar\right)}
  \right)\right\}
\ee
with the universal coefficient (cf.~\App{App_Cx})
\be
\gamma_0^{\rm spin}=\frac{6}{(4\pi)^2}-d_0\,,
\ee
and $\XRGIspin$ being again a renormalization group invariant matrix 
element, defined in the static effective theory.\footnote{
Alternatively, one can also construct a difference of squared energies
as the product $L^2\gamav\delgam$, which then behaves as
$\Cmag\XRGIspin\left(1+\rmO(1/z)\right)$ for $M\to\infty$, where $\Cmag$
is introduced in \App{App_Cx} as well.
}
The relative perturbative uncertainty of $\Cspin$ is $\rmO(\alpha^2)$,
since here the three-loop anomalous dimension is not known.
As illustrated in \fig{f_Cspin}, the difference between the one-loop
anomalous dimension and the two-loop one is tiny. 
Since this may, however, be accidental rather than representing the
behaviour of the series, we shall take the size of the three-loop term 
met in $\Cps$, \fig{f_Cpert}, as our uncertainty.
A better estimate of this uncertainty would require the knowledge of 
$\gamma_2^{\rm spin}$.

%
\FIGURE{
\epsfig{file=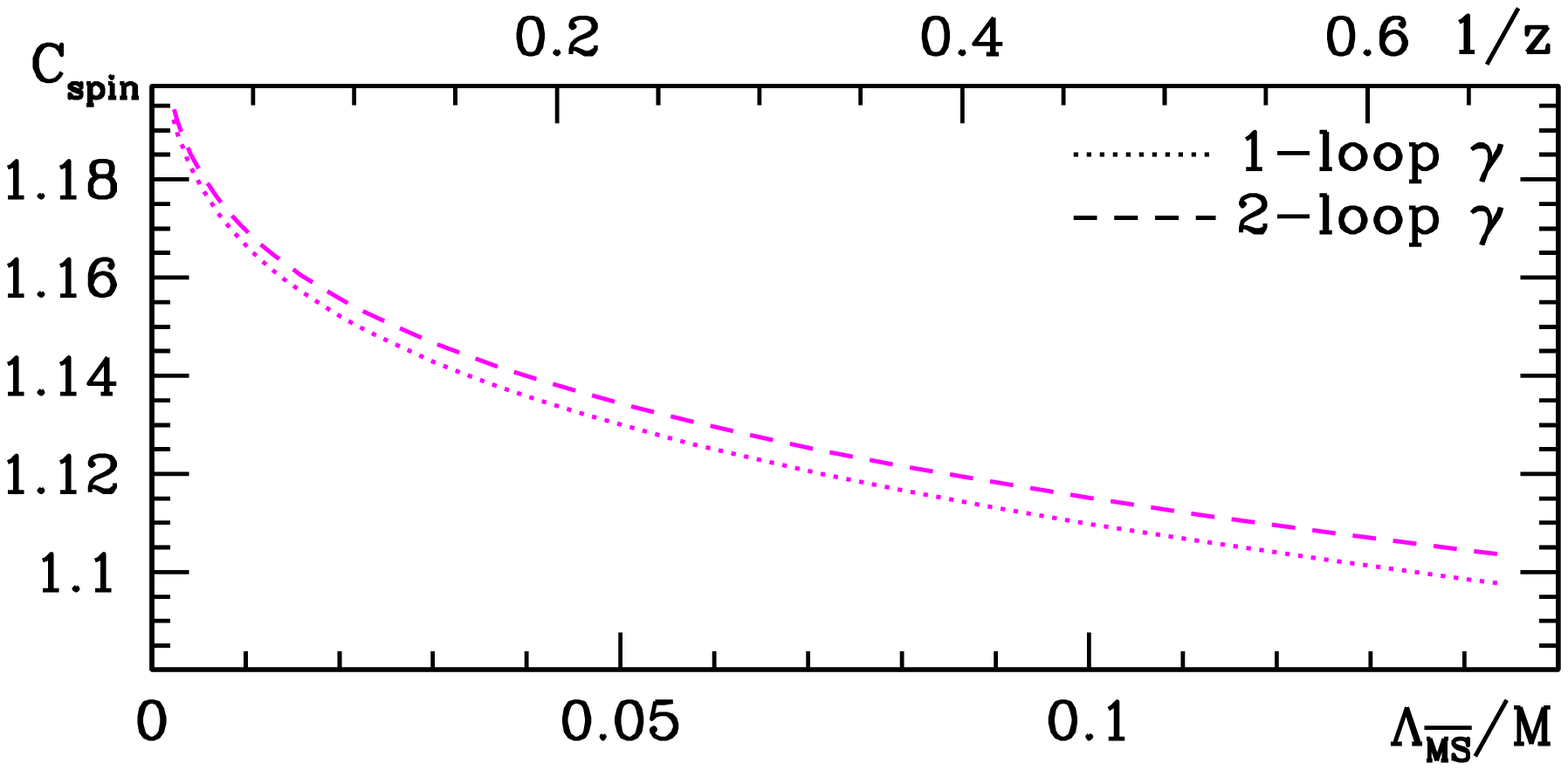,width=9cm} 
\vspace{-3.5cm}
\caption{The function $C_{\rm spin}$ evaluated in perturbation theory.
}\label{f_Cspin}
}
%

\section{Quantitative tests of the effective theory}
\label{Sec_res}
We have tested the predictions of the effective theory applied to the 
{\em quenched approximation} of QCD by evaluating the observables for 
one geometry, namely $T=L\equiv L_0$, and for $\theta=0.5$. 
The use of the quenched approximation should not be worrying in this 
context, since it is of course used both in the effective theory and 
in QCD. 
Furthermore, although we set the light quark mass to zero, $1/L_0$ 
provides an infrared cutoff and there are no singularities in the 
chiral limit. 

The numerical simulations are done on lattices with various resolutions 
$a/L_0$ followed by a continuum extrapolation. 
In all cases, $L_0$ is fixed by imposing
\bea \label{e_gbarl0}
  \gbar^2(L_0/2)=1.8811\,,
\eea
where $\gbar(L)$ is the renormalized coupling at length scale $L$ in the 
SF scheme \cite{alpha:SU3}. 
It is known that $L_0 \approx 0.2\,\fm$ \cite{msbar:pap1,HQET:pap1}. 
The (purely technical) reasons for the precise definition (\ref{e_gbarl0}) 
are detailed in \Ref{HQET:pap1}. 
Table~1 of \Ref{HQET:pap2} lists the bare coupling $g_0$ for each 
resolution $L_0/a$, and this reference also explains how the bare quark 
masses are fixed to ensure a massless light quark and a prescribed value
$M$ for the heavy quark.
\subsection{Results in the static approximation}
Some of the leading-order terms in the HQET expansions described in the 
previous section are known exactly due to the 
spin symmetry \cite{stat:symm1,stat:symm2}, but for two of the expansions 
we have computed the non-trivial leading order from a lattice simulation 
in static approximation. 

%
\FIGURE{
\epsfig{file=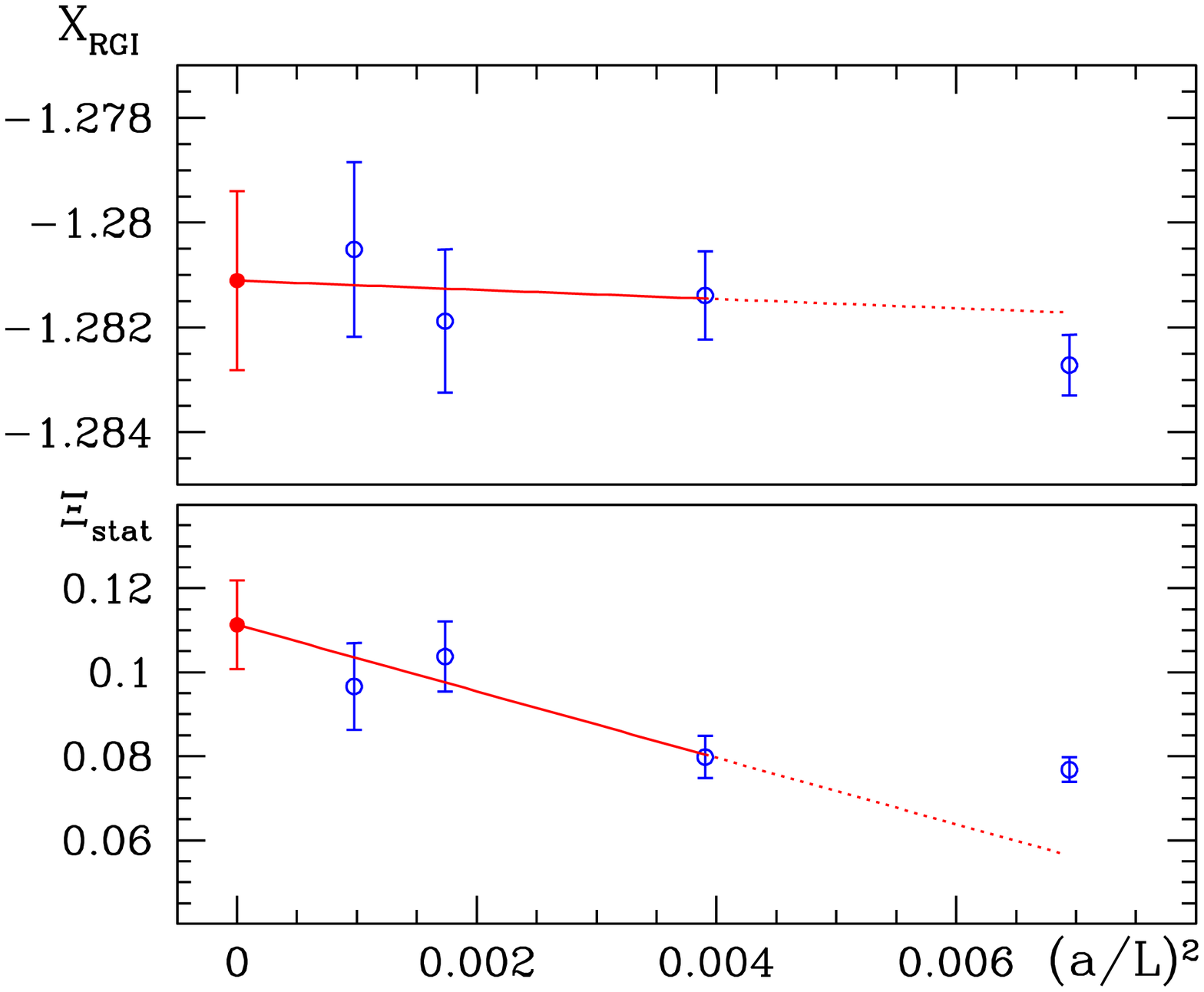,width=9cm} 
\vspace{0.5cm}
\caption{
Continuum limit extrapolations of $\XRGI(L_0)$ (top) and
$\gamdifstat(L_0)$ (bottom). 
The fits adopt simulation data generated with the HYP-link static quark 
action, see text.
The $a/L$ independent part of the error of $\XRGI$ is \emph{not} 
included in the graph.
}\label{fig:static_CL}
}
%
The first one is the matrix element of the time component of the axial 
current, $\XRGI(L)$ in \eq{e_XRGI}.
Indicating explicitly also the dependence on the bare coupling $g_0$ and 
the lattice spacing $a$, it is given by
\bea \label{e_XRGIlat}
 \XRGI(L_0) = \lim_{a/L_0 \to 0}\ZRGI(g_0)\,X(g_0,L_0/a)
 \,.
\eea
From the relation of $X$ to the very definition of the renormalization 
factor $\ZRGI$, as detailed in \Ref{zastat:pap3}, one obtains the 
explicit form 
\be 
\ZRGI\,X = 
\frac{\PhiRGI}{\Phi(\mu=1/L_0)}
\times
\left.\frac{\left(\fonehh\,f_1\right)^{1/4}}{\sqrt{\fonestat}}
\right|_{L=L_0}
\times \left[\,X\,\right]_{\,{\rm tree-level}}
\label{e_XRGIform}
\ee
for the right-hand side of \eq{e_XRGIlat}.  
All quantities that enter the above expression have been defined in the 
latter reference.
For its numerical evaluation we take the non-perturbatively $\Oa$ 
improved action of \cite{impr:pap3} for the light quarks and the improved 
discretizations of \cite{fbstat:pap1} for the heavy quark. 
We note in passing that, using the data of \cite{zastat:pap3}, we first 
evaluated this quantity with the Eichten-Hill action for the heavy 
quark \cite{stat:eichhill1}.
However, this resulted in an order of magnitude larger error for $\XRGI$ 
at $L_0/a=32$.

The limit $a/L_0\to0$ is taken by a linear fit in $(a/L_0)^2$ as 
illustrated in the upper diagram of \fig{fig:static_CL}, referring to the
data set from a simulation with the static quark action built from
HYP-links \cite{HYP:HK01,fbstat:pap1}.
We quote the result from a fit with $L_0/a\geq 16$ as our continuum 
result,
\bea \label{e_xrgicont}
   \XRGI(L_0) = -1.281(9) \,,
\eea
which also receives an error contribution from the (regularization 
independent) factor $\PhiRGI/\Phi(1/L_0)$ 
in \eq{e_XRGIform} \cite{zastat:pap3}. 
Including all points with $L_0/a\geq 12$ in the fit yields a compatible 
continuum value with smaller error. 
In the same way we obtain (cf.~\eq{e_gamdifstat} and the lower diagram 
of \fig{fig:static_CL}): 
\bea
  \gamdifstat(L_0) = 0.11(1) \,.
\eea
\subsection{Results at finite $M$ and comparison} 
%
\FIGURE{
\epsfig{file=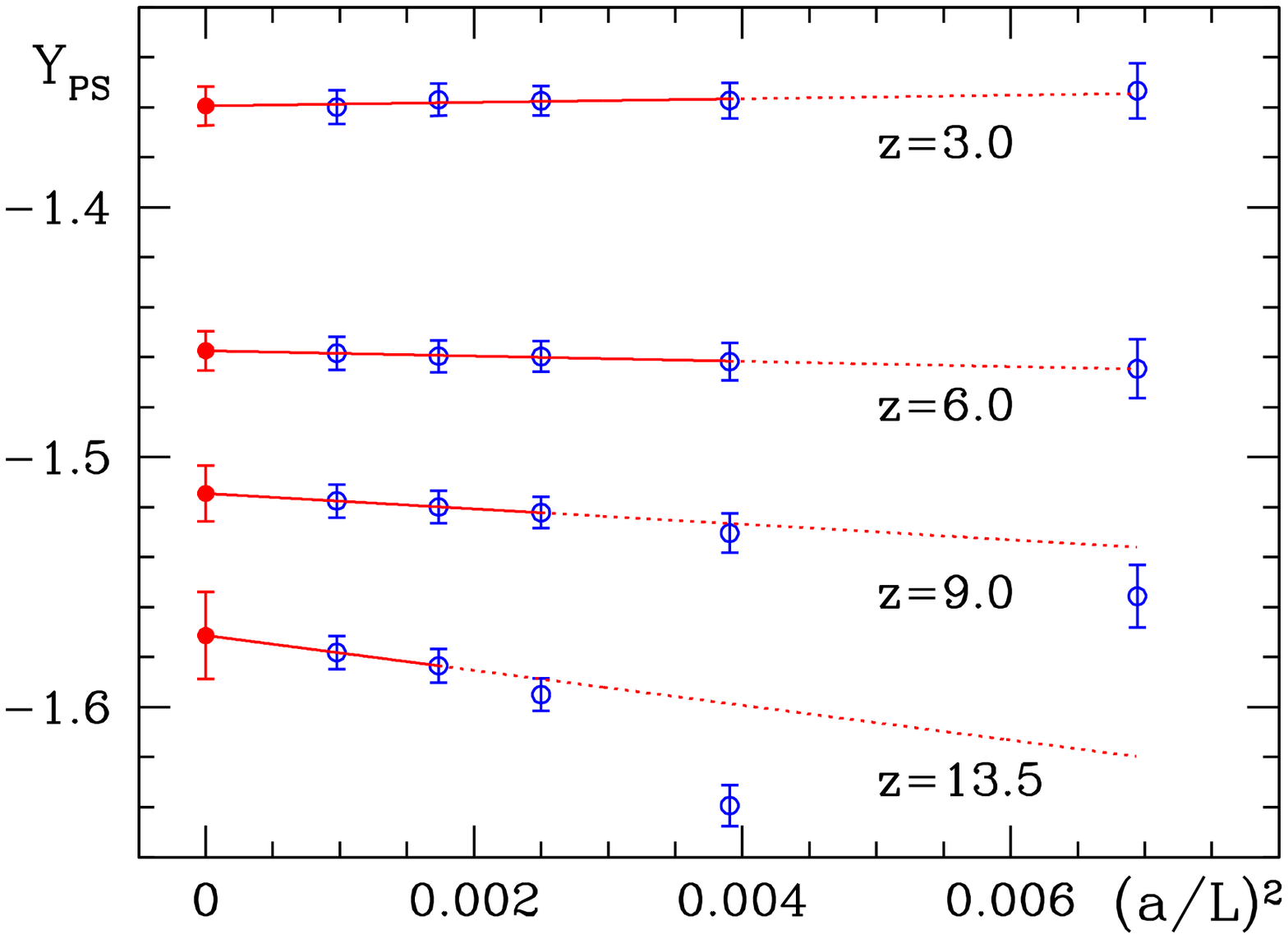,width=9cm} 
\caption{
Continuum extrapolations for some $z$-values spanning the entire range 
of $z$. The linear fit in $(a/L)^2$ is shown by the full line and
extended also to those values of $a/L$ that did not participate in the 
fit by a dotted line.
}\label{fig:yR_CL}
}
%
The finite-mass (quenched) QCD observables are obtained from similar 
extrapolations of lattice results at finite $a/L$.
However, as the variable $z$ is increased, the quark mass in lattice units 
grows at a given resolution $a/L$. 
A perturbative computation \cite{zastat:pap2} as well as our 
non-perturbative study suggest that $\Oa$ improvement may be trusted only 
below a certain value of the quark mass in lattice units. 
It is therefore necessary to impose a cut on the quark mass.  
For a given $z$, this cut translates into estimates of the coarsest 
lattice resolutions beyond which the lattice data are to be omitted from 
the continuum extrapolations.
As in \Ref{HQET:pap2} we impose $aM<0.6$ and follow this reference on all 
other details of the lattice computations as well as the extraction of the 
continuum limits.
For illustration we just show the continuum extrapolations of 
$\Yr(L,M)$, $L=L_0$, at selected values of $z$ in \fig{fig:yR_CL}. 
In addition we mention a few features equally true for the continuum limit 
extrapolations of the other observables, which enter our investigation but 
are not shown in figures. 
\begin{itemize}
\item The slopes in $a/L$ are rather small. 
\item The error in the continuum limit grows with $z$, because less 
      lattices can be used in the extrapolation at large $z$. 
\item The continuum limit is compatible with the values at the smallest 
      two lattice spacings. 
      Its error is conservative.
\end{itemize}
In \App{App_numres} we collect our numerical results both at finite 
lattice spacing and in the continuum limit.

%
\TABLE{
\begin{tabular}{rcr@{.}lr@{.}lcr@{.}lr@{.}lr@{.}l}
\hline\hline \\[-2.0ex]
&& \multicolumn{4}{c}{Linear}
&& \multicolumn{6}{c}{Quadratic} \\[1.0ex]
   Quantity
&& \multicolumn{2}{c}{$a_0$} & \multicolumn{2}{c}{$a_1$}
&& \multicolumn{2}{c}{$a_0$} & \multicolumn{2}{c}{$a_1$} 
&  \multicolumn{2}{c}{$a_2$} \\[1.0ex]
\hline\hline \\[-2.0ex]
\multicolumn{13}{c}{$z$-range: $3.0$--$13.5$}  \\[1.0ex]
\hline \\[-2.0ex]
   $\Yr/\Cps$
&& \multicolumn{2}{c}{ } & \multicolumn{2}{c}{ }
&& $-1$&$281(7)$ & $0$&$64(8)$  & $-1$&$0(2)$  \\
   $\Yv/\Cv$
&& \multicolumn{2}{c}{ } & \multicolumn{2}{c}{ }
&& $-1$&$281(7)$ & $-0$&$63(9)$ & $0$&$3(2)$   \\
   $\Rr/\Cpsv$
&& \multicolumn{2}{c}{ } & \multicolumn{2}{c}{ }
&& $1$&$0$       & $-0$&$89(1)$ & $1$&$06(3)$  \\
   $L\gamav/(z\,\Cmass)$
&& \multicolumn{2}{c}{ } & \multicolumn{2}{c}{ }
&& $1$&$0$       & $0$&$42(3)$  & $0$&$14(10)$ \\
   $L\delgam/\Cspin$
&& \multicolumn{2}{c}{ } & \multicolumn{2}{c}{ }
&& $0$&$0$       & $-1$&$69(6)$ & $0$&$8(2)$   \\
   $\gamdif$
&& \multicolumn{2}{c}{ } & \multicolumn{2}{c}{ }
&& $0$&$109(8)$  & $0$&$7(1)$   & $-0$&$8(3)$  \\[1.0ex]
\hline \\[-2.0ex]
\multicolumn{13}{c}{$z$-range: $5.15$--$13.5$} \\[1.0ex]
\hline \\[-2.0ex]
   $\Yr/\Cps$
&& $-1$&$277(7)$ & $0$&$45(4)$  
&& $-1$&$281(8)$ & $0$&$7(1)$   & $-1$&$3(5)$  \\
   $\Yv/\Cv$
&& $-1$&$281(9)$ & $-0$&$59(6)$  
&& $-1$&$281(7)$ & $-0$&$6(1)$  & $0$&$2(6)$   \\
   $\Rr/\Cpsv$
&& $1$&$0$       & $-0$&$722(7)$  
&& $1$&$0$       & $-0$&$91(1)$ & $1$&$18(5)$  \\
   $L\gamav/(z\,\Cmass)$
&& $1$&$0$       & $0$&$44(2)$  
&& $1$&$0$       & $0$&$41(6)$  & $0$&$2(3)$   \\
   $L\delgam/\Cspin$
&& $0$&$0$       & $-1$&$56(4)$ 
&& $0$&$0$       & $-1$&$62(6)$ & $0$&$4(2)$   \\
   $\gamdif$
&& $0$&$112(8)$  & $0$&$54(7)$ 
&& $0$&$110(9)$  & $0$&$6(2)$   & $-0$&$5(7)$  \\[1.0ex]
\hline\hline \\[-2.0ex]
\end{tabular}
\caption{
Fit parameters describing the $z$-dependences of the form
$a_0+a_1/z+a_2/z^2$. 
Where the constant term, $a_0$, is listed without an error, it is 
constrained to the prediction of the static effective theory.
(In case of the entire $z$-range, only the quadratic fit results are 
given.)
All these fits have an acceptable goodness-of-fit.
}\label{tab:fitres}

}
%
\FIGURE{
\epsfig{file=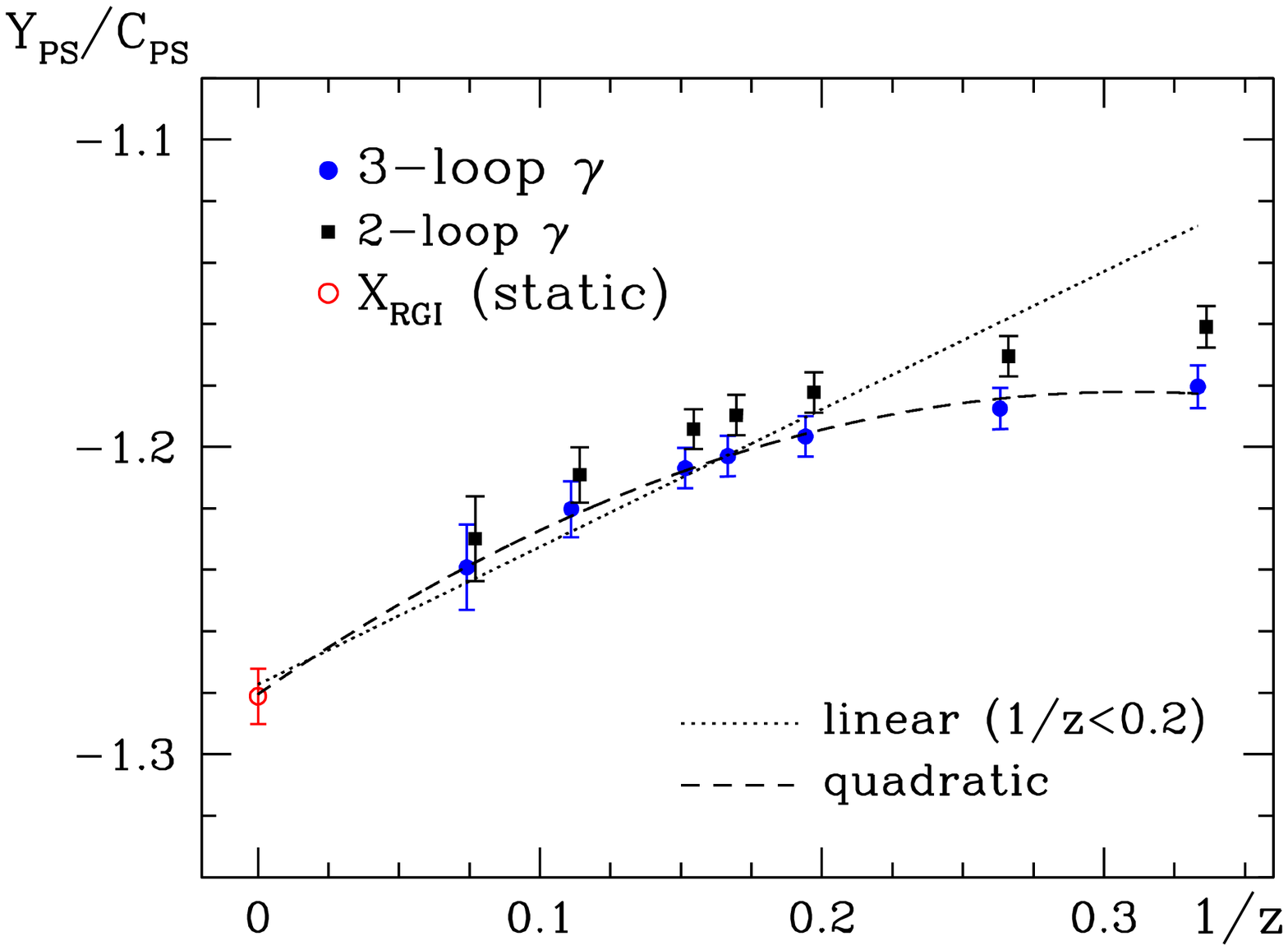,width=9cm} 
\caption{
The ratio $\Yr(L,M)/\Cps(M/\lMSbar)$ with $\Cps$ using the 
two- and the three-loop anomalous dimension of the static axial current. 
Error bars do not contain the perturbative uncertainty in $\Cps$.
The fits shown refer to the three-loop evaluation of $\Cps$ and include
the result for $\XRGI$ in the static limit. 
(Points using only the two-loop anomalous dimension are slightly displaced
here.)
}\label{fig:yR_z}
}
%
\FIGURE{
\epsfig{file=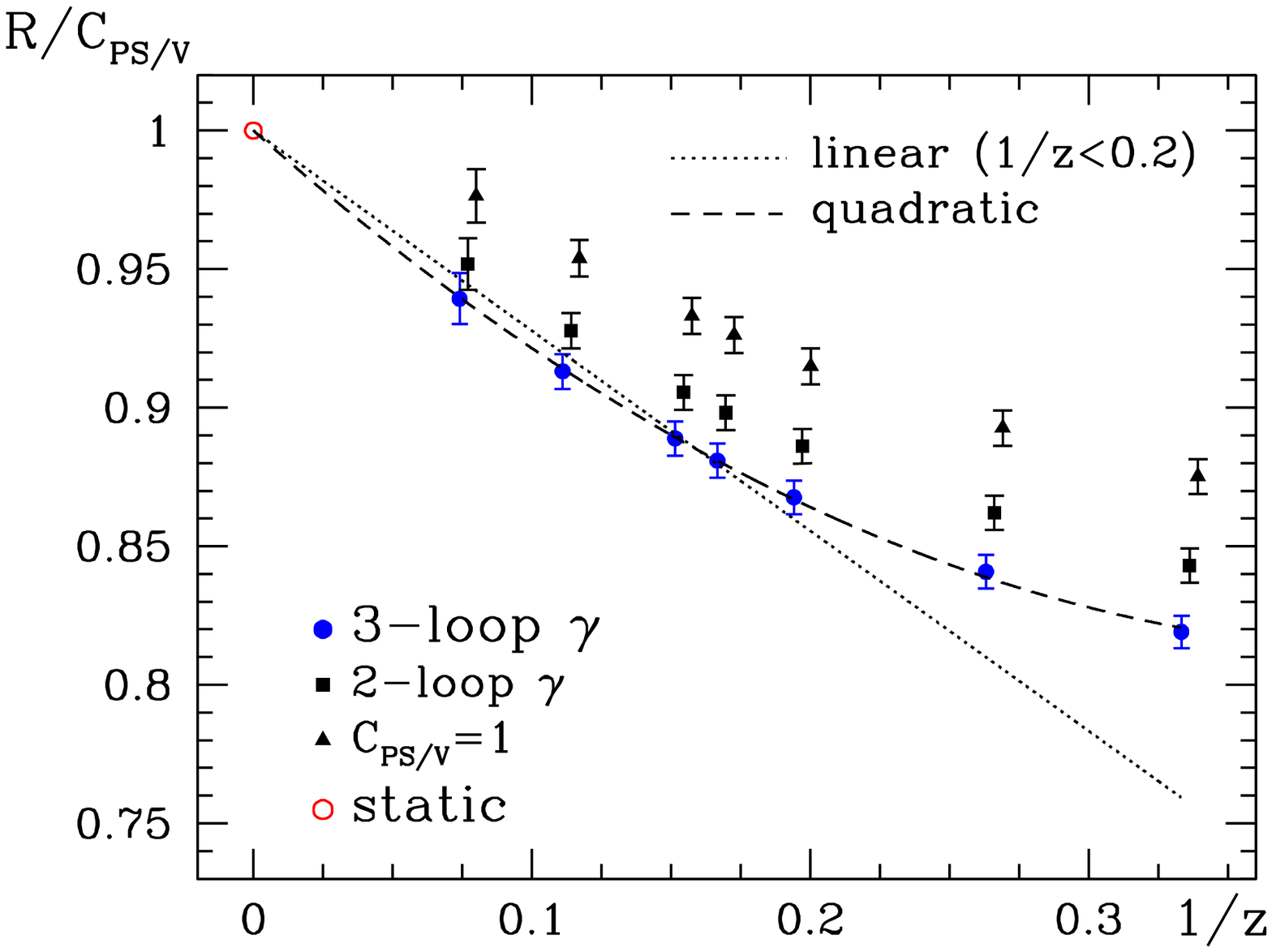,width=9cm} 
\caption{
As in \fig{fig:yR_z}.
The fits use the evaluation of $\Cpsv$ with the two-loop
matching coefficient between HQET and QCD; they are constrained to the 
prediction of the static effective theory.
(On the level of perturbative orders, two-loop matching belongs to the 
three-loop anomalous dimension of the currents, cf.~\App{App_Cx}). 
}\label{fig:rR_z}
}
%

We are now ready to compare the continuum results with the predictions 
of HQET. 
Before that, we remind the reader that energies of order $2/L$ still 
contribute significantly to our observables. 
It is thus possible that the $1/M$-expansion breaks down earlier in our 
finite-volume situation than it does in large volume. 
However, our numerical results do not give any indication of such a 
behaviour.

Let us start with the current matrix elements.
The prediction for the matrix element of the axial current is 
$\Yr(L,M)/\Cps = \XRGI(L) +\rmO(1/z)$.
In this combination, plotted in \fig{fig:yR_z}, the perturbatively 
computed coefficient $\Cps(M/\lMSbar)$ compensates a significant part of 
the mass dependence of $\Yr$. 
The finite-mass $\Yr/\Cps$ is obviously well compatible with approaching 
the static result, \eq{e_xrgicont}, as $1/z \to 0$. 
To quantify the deviations from the static limit at finite $z$, we fit all 
data points, including $1/z=0$, to first- and second-order polynomials 
in $1/z$:
\bes \label{e_yrfit}
  \frac{\Yr}{\Cps} \,=\, a_0 \,+\, \frac{a_1}{z} \,+\, \ldots\,\,.
\ees
In these fits also the uncertainty in $\Cps$ is taken into account
(i.e.~half of the difference between $\Cps$ evaluated with the two-loop and 
three-loop anomalous dimension). 
We perform them separately to the data in the range $1/z<0.2$, which means 
masses around the b-quark mass and higher \cite{HQET:pap1}, and over the 
whole range, see \tab{tab:fitres}. 
Comparing $a_1$ obtained from the quadratic fit over the whole range 
($z=3.0$--$13.5$) and the linear fit for $1/z<0.2$ ($z\geq 5.15$), the 
change is not so small. 
This indicates that a precise identification of the first-order correction 
is not possible with our precision and range of $z$. 
However, the rough magnitude of $a_1 \approx 0.6$ can be inferred and, 
more importantly, it is clear that the overall magnitude of $
1/z$-corrections is reasonably small.
It is also relevant to remember that \eq{e_yrfit} is only an approximate 
parametrization of the $z$-dependence, since the renormalization of the 
higher-dimensional operators in the effective theory will introduce 
logarithmic modifications of the simple power series. 
It is thus conceivable that these logarithms account for some of the 
curvature seen in \fig{fig:yR_z}.

%
\FIGURE{
\epsfig{file=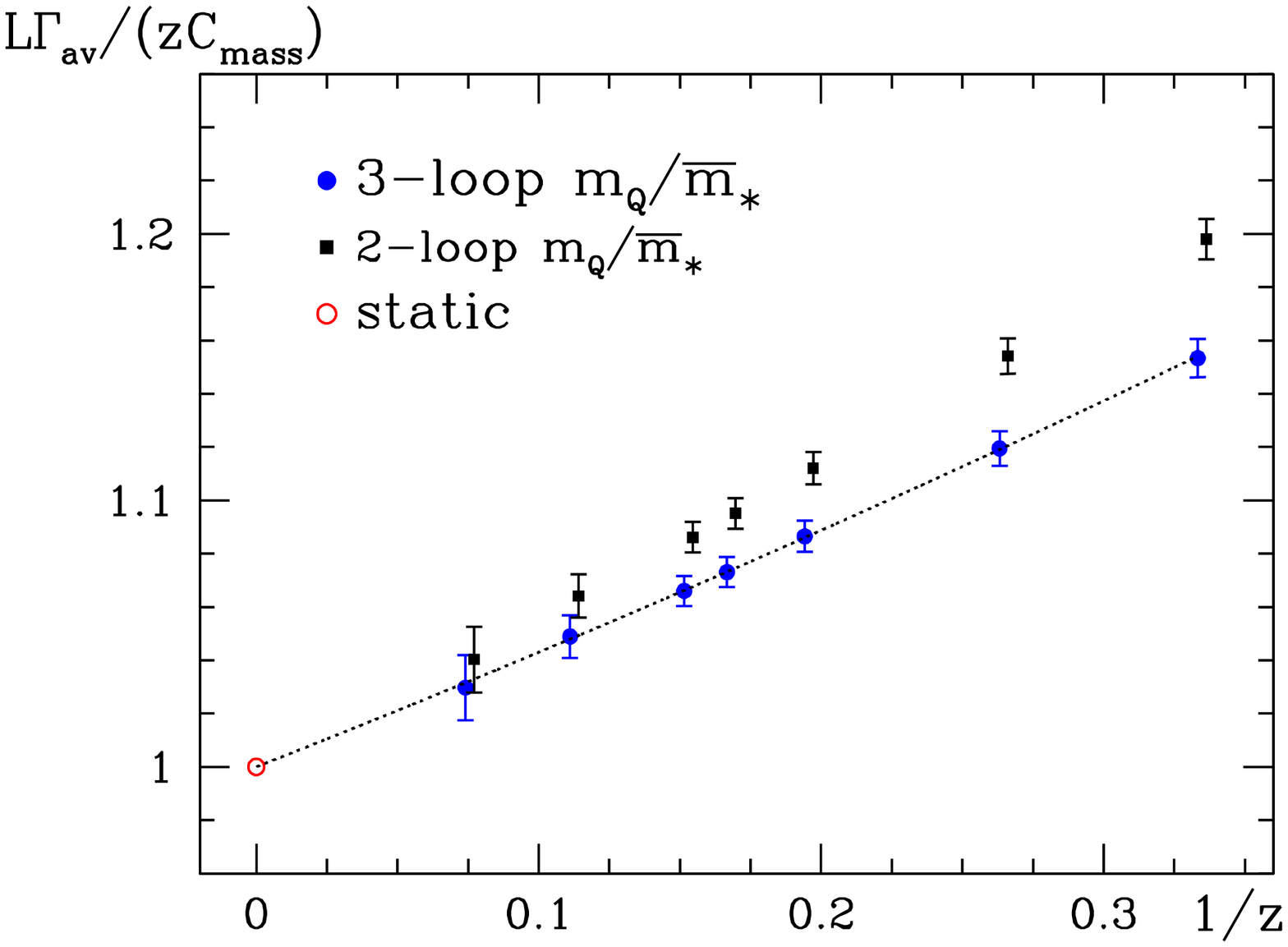,width=9cm} 
\caption{
Quadratic fit over the entire $z$-range of the combination 
$L\gamav/(z\,\Cmass)$, which is constrained to approach 1 in the static 
limit.
It employs the three-loop relation between the pole and the $\MS$ mass;
the results using this relation to only two loops are also depicted 
for comparison.
}\label{fig:gamav_z}
}
%
\FIGURE{
\epsfig{file=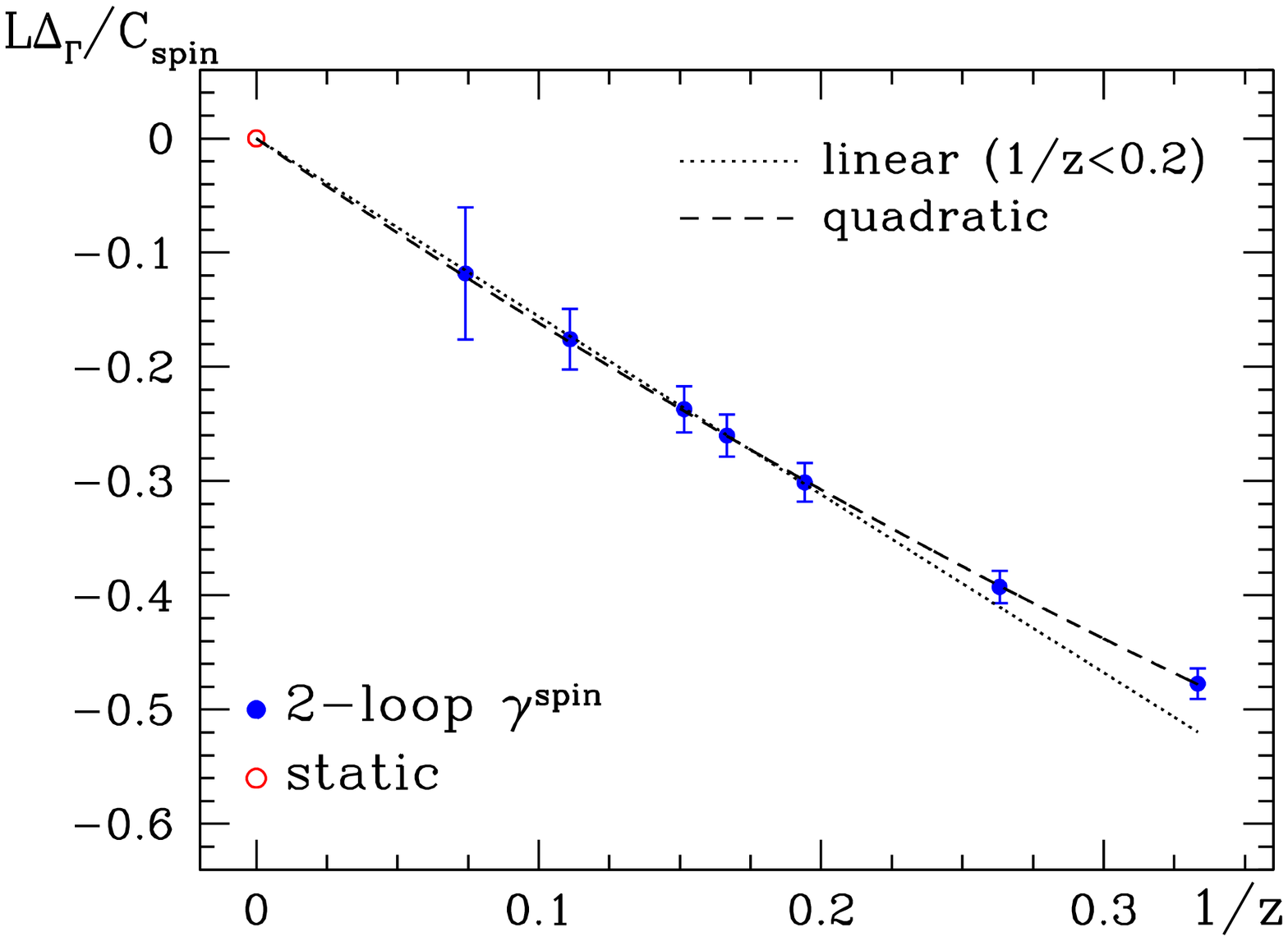,width=9cm} 
\caption{
Fits in $1/z$ constrained to the static result and referring to the
two-loop evaluation of $\Cspin$.
The linear fit is only based on the heavier quark mass points with 
$z=5.15$--$13.5$, while the quadratic one (i.e.~also allowing for a 
$1/z^2$-term) includes all points.
}\label{fig:delgam_z}
}
%

For the ratio of matrix elements $\Rr(L,M)$, \eq{rR2stat}, the lowest-order 
term in the $1/z$-expansion is fixed to be 1 by the spin symmetry of the 
static theory. 
As is reflected by \fig{fig:rR_z}, the results at finite $1/z$ are well 
compatible with this, if the function $\Cpsv$ is evaluated including at 
least the two-loop anomalous dimensions.
Fits to $\Rr/\Cpsv$ are performed in complete analogy to \eq{e_yrfit}. 
The corresponding parameters of \tab{tab:fitres} are again of order unity. 
In that table we also include the parameters of the analogous fits to the 
quantity $\Yv/\Cv$. 

Turning our attention to the effective energies, introduced in 
\sect{s_corr}, a first rough test of HQET is the behaviour of the 
spin-averaged energy $\gamav$.
\Fig{fig:gamav_z} confirms the expectation that the combination 
$L\gamav/(z\,\Cmass)$ approaches 1 up to $1/z$-corrections as $1/z\to 0$ 
(see~\eq{e_gamasympt}).
Note that the $(1/z)^2$-corrections in $\gamav$ are very small, which is 
of particular interest for the computation of the b-quark mass in static 
approximation \cite{HQET:pap1,HQET:pap2}.
There the quantity $\gamav$ was used in order to non-perturbatively match 
the quark mass of the effective theory to the one in QCD.
The dominant error in the final estimate for the quark mass, $M_{\rm b}$, 
is expected to originate from this matching and is of order
$M_{\rm b} \times (1/z)^2$. From the values of $a_2$ 
and $z_{\rm b}\equiv\mb L \approx 5$, we may estimate the relative error 
to be roughly of the order of $0.2 \times (1/5)^2 \approx 1\%$.
As seen from the figure, this conclusion is not very much affected either 
by the perturbative uncertainty visible in the graph as a difference 
between the two-loop and three-loop approximations for the pole mass 
$m_Q$ in \eq{e_cmass}; 
the $1/z^2$-curvature is not very different.

The spin splitting $\delgam(L,M)$, which vanishes in the static limit, 
is displayed in \fig{fig:delgam_z}.
It is in good agreement with the HQET prediction but exhibits a rather 
large $1/z$-coefficient.\footnote{
We note that the slope $\XRGIspin(L)$ is computable in the effective 
theory in a very similar way to $\XRGI$, because the associated operator 
$\heavyb\vecsigma\cdot\vecB\,\heavy$ does not mix with any other; 
its renormalization may be computed non-perturbatively using the methods 
of \Ref{zastat:pap3}. 
The comparison of the result to the data at finite mass would presumably 
be limited in precision by the present perturbative uncertainty in 
$\Cspin$. 
This limitation is likely to apply to the (large-volume) mass splitting 
between the ${\rm B}^*$- and the ${\rm B}$-meson as well. 
}

Finally, we successfully test \eq{e_gamdif} in \fig{fig:gamdif_z}. 
Recall that a simple kinematical consideration leads one to expect 
the $1/z$-expansion to only be applicable at smaller values of $1/z$ for
this observable (cf.~\sect{s_ener}). 
On the other hand, reparametrization invariance allows to exclude
logarithmic modifications of the $1/z$-term;
in contrast to our other tests, \fig{fig:gamdif_z} does not involve any
perturbative conversion factor.
%
\FIGURE{
\epsfig{file=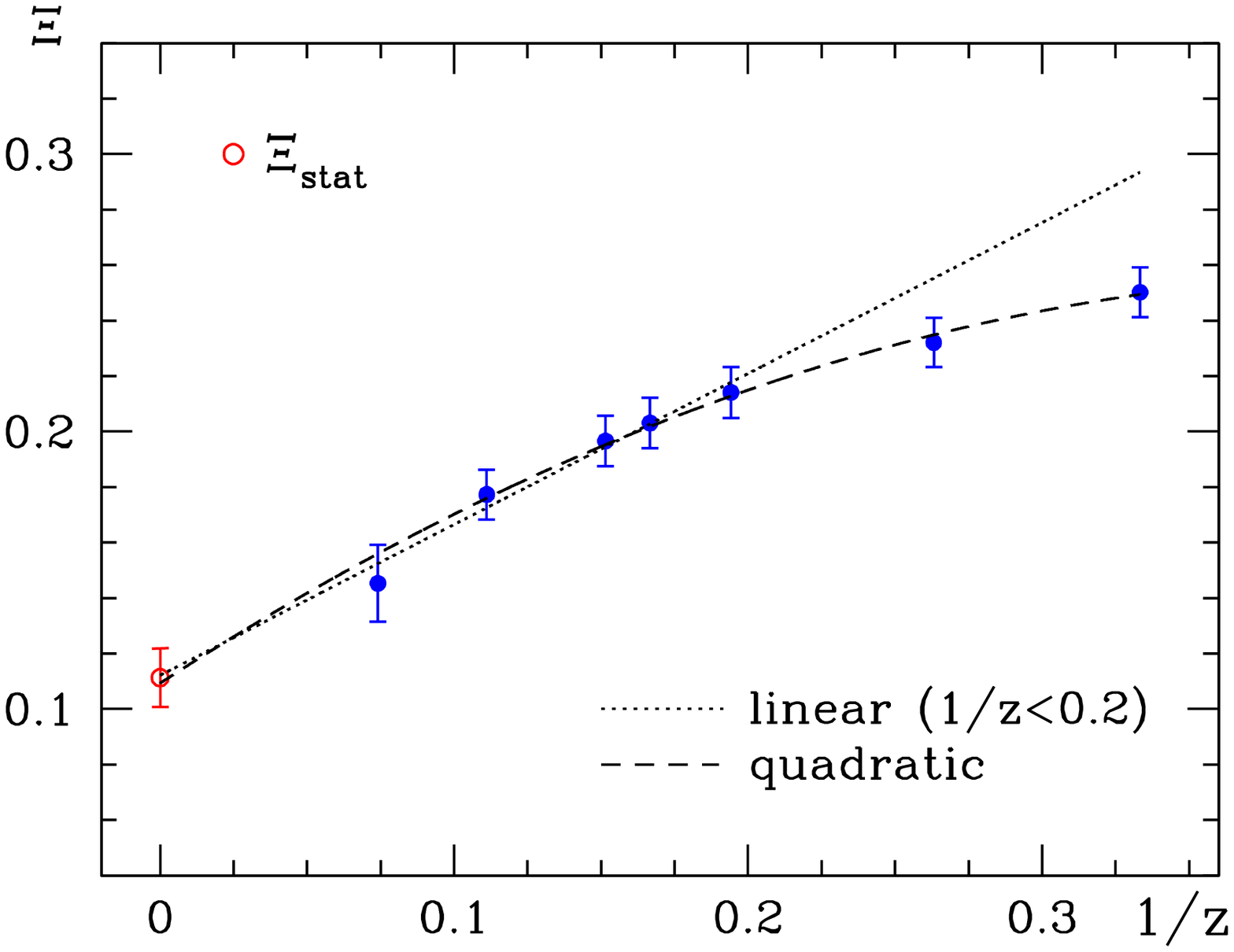,width=9cm} 
\caption{
Linear and quadratic fits of the observable $\gamdif$ as 
in \figs{fig:yR_z}--\ref{fig:delgam_z}, where the static data point, 
$\gamdifstat$, is included in the fits.
}\label{fig:gamdif_z}
}
%

\section{Conclusions}
\label{Sec_concl}
All of the comparisons of QCD observables with the predictions of
HQET discussed in this work represent tests of the effective theory, 
which it passed successfully. 
As a significant improvement of earlier non-perturbative (lattice) tests, 
these comparisons are performed after first taking the 
{\em continuum limit} of the {\em non-perturbatively renormalized}
quantities. 
Thus, worries that $\rmO(a)$ effects, in particular those that grow with 
the quark mass, may afflict the large-mass behaviour in QCD are removed.

For a precise judgement of \figs{fig:yR_z}--\ref{fig:gamdif_z} it is 
important to be aware of the level of precision of renormalization 
and matching. 
The renormalization problem of the static axial current has been solved 
non-perturbatively in \cite{zastat:pap3} and with this information all 
points at $1/z=0$ are known without any residual perturbative uncertainty. 
For the quantity $\gamdif$ shown in \fig{fig:gamdif_z}, this is also true 
at finite $1/z$. 
However, in general we need to know the functions $C_{\rm X}$, which 
relate the QCD observables at finite mass $M$ (and thus finite $z=ML$) to 
the renormalization group invariants of the effective theory, such as 
$\XRGI$ in \fig{fig:yR_z}. 
While the latter are unambiguously defined and have been computed 
non-perturbatively, the former are known only in perturbation theory and 
have errors of order $\alpha^n$, with $\alpha$ evaluated at the scale of 
the heavy quark mass. 
We have discussed these errors. 
On a phenomenological level they are under control due to the computations
\cite{BroadhGrozin2,ChetGrozin}, except for $\Cspin$, where a 
next-to-next-to-leading-order computation is not yet available. 
Still, one should remember that -- strictly speaking -- the isolation of 
$1/m$-corrections by looking at the difference to the leading-order HQET 
result is only possible when $C_{\rm X}$ is known non-perturbatively:
parametrically, $\rmO([\alpha(m)]^n)$ corrections are always larger 
than  $\rmO(1/m)$. 
Nevertheless, our results in \figs{fig:yR_z}--\ref{fig:gamdif_z} are 
compatible with the $z^{-n}$ power corrections dominating over the 
perturbative ones in the accessible range of $z$. 
Fitting them by a simple polynomial in $1/z$, the coefficients turn out 
to be of order 1 as naively expected. 
Therefore, at the b-quark mass, which corresponds to 
$z_{\rm b}\equiv\mb L \approx 5$ for our value of 
$L=L_0 \approx 0.2\,\fm$, the effective theory is very useful.

Beyond the general interest of providing a non-perturbative test for HQET,
these results are important for the programme of \Ref{HQET:pap1}. 
There it was suggested to determine the $1/m_{\rm b}$-corrections to 
B-physics matrix elements from a simulation of HQET on the lattice.
The coefficients $c_k$ of the various terms in the HQET Lagrangian
and of the effective composite fields are to be determined by matching
HQET and QCD in a finite volume, and it was proposed to employ a value of 
$L$ similar to $L_0$ in this step. 
This value has to be large enough such that (i) $z_{\rm b}$ is in the 
range where HQET is applicable with small corrections, yet (ii) $L_0$ has 
to be small enough to allow for the computation of the QCD observables at 
$m=m_{\rm b}$ with small $a$-effects. From \tab{tab:fitres} we conclude 
that $L=L_0$ is indeed promising.
In fact, in the previous section we roughly estimated that the correction
to the static limit computation of the RGI b-quark mass 
\cite{lat01:mbstat,HQET:pap1} is only of the order of $1\%$. 

However, this application appears to be a particularly favourable case and 
it would be of advantage to have a larger value of $z_{\rm b}$ 
(i.e.~larger $L$) in the matching step to suppress higher-order terms. 
Our analysis suggests that this is indeed possible, since we may 
determine QCD observables rather precisely in, say, the entire range 
$z\geq5$ by combining QCD results at finite $z$ with the static limit. 
Choosing, for instance, $L\approx2L_0$ in the matching step, one needs
the QCD observables for $z\approx 10$.
The errors of the quadratic fit functions evaluated at $z=\zb=10$ are 
typically 30\%--50\% smaller than the errors of the neighbouring points 
seen in \figs{fig:yR_z}--\ref{fig:gamdif_z}. 
Proceeding in this way, namely taking the fit functions as 
representations of the QCD observables in finite volume (of course within 
their errors), one may obtain the HQET parameters as functions of the 
quark mass and infer predictions for all quark masses larger than the 
minimal one considered.   
With the entire matching done non-perturbatively, the final HQET results 
will differ from QCD by $\rmO(1/m^2)$ if $1/m$-terms are properly 
included. No perturbative errors as in $C_{\rm X}$ enter in this 
programme.

We finally remind the reader that in the quantities discussed in this 
paper, as well as in the matching step just described, we are dealing 
with matrix elements of a mixture of energy eigenstates where states 
with energy of $\rmO(1/L)$ contribute.
Hence the $1/m$-expansion for the large-volume B-physics matrix elements 
may be expected to behave even better and should be well under control, 
once this is the case for the matching step.
\acknowledgments
We would like to thank M.~Della~Morte, T.~Mannel, T.~Feldmann and
N.~Tantalo for useful discussions.
This work is part of the ALPHA Collaboration research programme.
The largest part of the numerical simulations has been performed on the
APEmille computers at DESY Zeu\-then, and we thank DESY for allocating 
computer time to this project as well as the staff of the computer center 
at Zeuthen for their support.
In addition we ran a C-code based on the MILC Collaboration's public 
lattice gauge theory code \cite{code:MILC} on the PC cluster of the 
University of M\"unster.
This work is supported in part by the EU IHP Network on 
\emph{Hadron Phenomenology from Lattice QCD} 
under grant HPRN-CT-2000-00145 and by the Deutsche Forschungsgemeinschaft 
in the SFB/TR 09.
\clearpage

\begin{appendix}
\section{Results at finite lattice spacing}
\label{App_numres}
For any details on the lattice simulations and the subsequent analysis of 
the numerical data, the reader is referred to \cite{HQET:pap2} and 
references therein.
In \tab{tab:latres} we therefore directly list the results on the 
observables studied in this work at finite values of the lattice spacing 
as well as in the continuum limit.
As already mentioned in \sect{Sec_res}, the latter have been extracted by
linear extrapolations $(a/L)^2\rightarrow 0$ of the $\Oa$ improved lattice 
data following the same procedure as adopted in \Ref{HQET:pap2}.
\vfill
%
\TABLE{
\begin{tabular}{ccr@{.}lr@{.}lr@{.}lr@{.}lr@{.}l}
\hline\hline
   $L/a$ 
&& \multicolumn{2}{c}{$\Yr(L,M)$} 
&  \multicolumn{2}{c}{$\Yv(L,M)$}
&  \multicolumn{2}{c}{$\Rr(L,M)$}
&  \multicolumn{2}{c}{$L\delgam(L,M)$} 
&  \multicolumn{2}{c}{$\gamdif(L,M)$} \\
\hline\hline
\multicolumn{12}{c}{$z=3.0$}     \\
\hline
   $12$
&& $-1$&$35(1)$  & $-1$&$581(7)$ & $0$&$869(8)$  
&  $-0$&$509(10)$ & $0$&$245(4)$ \\
   $16$
&& $-1$&$357(7)$ & $-1$&$576(7)$ & $0$&$874(6)$  
&  $-0$&$512(10)$ & $0$&$256(5)$ \\
   $20$
&& $-1$&$357(6)$ & $-1$&$575(7)$ & $0$&$875(5)$  
&  $-0$&$521(11)$ & $0$&$259(6)$ \\
   $24$
&& $-1$&$357(6)$ & $-1$&$576(7)$ & $0$&$874(5)$  
&  $-0$&$522(13)$ & $0$&$259(8)$ \\
   $32$
&& $-1$&$360(7)$ & $-1$&$577(7)$ & $0$&$875(5)$  
&  $-0$&$533(14)$ & $0$&$246(9)$ \\
   {\it CL}
&& ${\it -1}$&${\it 359(8)}$ & ${\it -1}$&${\it 576(8)}$ 
&  ${\it 0}$&${\it 875(6)}$ & ${\it -0}$&${\it 537(15)}$ 
&  ${\it 0}$&${\it 250(9)}$      \\
\hline
\multicolumn{12}{c}{$z=3.8$}     \\
\hline
   $12$
&& $-1$&$39(1)$  & $-1$&$591(7)$ & $0$&$884(8)$  
&  $-0$&$419(13)$ & $0$&$226(4)$ \\
   $16$
&& $-1$&$392(7)$ & $-1$&$584(7)$ & $0$&$890(6)$  
&  $-0$&$423(12)$ & $0$&$240(5)$ \\
   $20$
&& $-1$&$391(6)$ & $-1$&$582(7)$ & $0$&$892(5)$  
&  $-0$&$430(12)$ & $0$&$240(6)$ \\
   $24$
&& $-1$&$391(6)$ & $-1$&$582(7)$ & $0$&$891(5)$  
&  $-0$&$431(13)$ & $0$&$241(8)$ \\
   $32$
&& $-1$&$394(7)$ & $-1$&$583(7)$ & $0$&$892(5)$  
&  $-0$&$441(15)$ & $0$&$228(9)$ \\
   {\it CL}
&& ${\it -1}$&${\it 394(8)}$ & ${\it -1}$&${\it 581(8)}$ 
&  ${\it 0}$&${\it 893(6)}$ & ${\it -0}$&${\it 444(16)}$ 
&  ${\it 0}$&${\it 232(9)}$      \\
\hline
\multicolumn{12}{c}{$z=5.15$}    \\
\hline
   $12$
&& $-1$&$44(1)$  & $-1$&$609(7)$ & $0$&$904(8)$  
&  $-0$&$315(17)$ & $0$&$200(4)$ \\
   $16$
&& $-1$&$438(7)$ & $-1$&$596(7)$ & $0$&$910(6)$  
&  $-0$&$320(15)$ & $0$&$215(5)$ \\
   $20$
&& $-1$&$437(6)$ & $-1$&$592(7)$ & $0$&$913(5)$  
&  $-0$&$328(15)$ & $0$&$216(6)$ \\
   $24$
&& $-1$&$437(6)$ & $-1$&$591(7)$ & $0$&$913(5)$  
&  $-0$&$327(16)$ & $0$&$218(8)$ \\
   $32$
&& $-1$&$436(7)$ & $-1$&$589(7)$ & $0$&$914(5)$  
&  $-0$&$341(17)$ & $0$&$210(9)$ \\
   {\it CL}
&& ${\it -1}$&${\it 435(8)}$ & ${\it -1}$&${\it 586(8)}$ 
&  ${\it 0}$&${\it 915(6)}$ & ${\it -0}$&${\it 343(19)}$ 
&  ${\it 0}$&${\it 214(9)}$      \\
\hline\hline
\end{tabular}
\caption{
Lattice results on the quantities of this work for different values of the
dimensionless renormalization group invariant heavy quark mass, $z=ML$,
with $L\equiv L_0\approx 0.2\,\Fm$.
The full sets of simulation parameters can be inferred from tables~1 and~4
of \Ref{HQET:pap2}, where also the results for
$L\gamav$ were published.
The quoted errors cover the statistical as well as the systematic 
uncertainties, including those originating from the fact that in the 
numerical simulations $z$ can only be fixed within some finite precision.
Continuum limits are displayed in italics. 
}\label{tab:latres}

}
%
\vfill
%
\TABLE{
\begin{tabular}{ccr@{.}lr@{.}lr@{.}lr@{.}lr@{.}l}
\hline\hline
   $L/a$ 
&& \multicolumn{2}{c}{$\Yr(L,M)$} 
&  \multicolumn{2}{c}{$\Yv(L,M)$}
&  \multicolumn{2}{c}{$\Rr(L,M)$}
&  \multicolumn{2}{c}{$L\delgam(L,M)$} 
&  \multicolumn{2}{c}{$\gamdif(L,M)$} \\
\hline\hline
\multicolumn{12}{c}{$z=6.0$}     \\
\hline
   $12$
&& $-1$&$46(1)$  & $-1$&$620(7)$ & $0$&$913(8)$  
&  $-0$&$270(20)$ & $0$&$186(4)$ \\
   $16$
&& $-1$&$462(7)$ & $-1$&$604(7)$ & $0$&$920(6)$  
&  $-0$&$276(17)$ & $0$&$203(5)$ \\
   $20$
&& $-1$&$460(6)$ & $-1$&$597(7)$ & $0$&$923(5)$  
&  $-0$&$283(17)$ & $0$&$205(6)$ \\
   $24$
&& $-1$&$460(6)$ & $-1$&$595(7)$ & $0$&$924(5)$  
&  $-0$&$282(17)$ & $0$&$207(8)$ \\
   $32$
&& $-1$&$458(7)$ & $-1$&$593(7)$ & $0$&$924(5)$  
&  $-0$&$296(19)$ & $0$&$198(9)$ \\
   {\it CL}
&& ${\it -1}$&${\it 457(8)}$ & ${\it -1}$&${\it 589(8)}$ 
&  ${\it 0}$&${\it 926(6)}$ & ${\it -0}$&${\it 297(21)}$ 
&  ${\it 0}$&${\it 203(9)}$      \\
\hline
\multicolumn{12}{c}{$z=6.6$}     \\
\hline
   $12$
&& $-1$&$48(1)$  & $-1$&$629(7)$ & $0$&$918(8)$  
&  $-0$&$243(21)$ & $0$&$178(4)$ \\
   $16$
&& $-1$&$477(8)$ & $-1$&$609(7)$ & $0$&$926(6)$  
&  $-0$&$251(19)$ & $0$&$196(5)$ \\
   $20$
&& $-1$&$474(6)$ & $-1$&$601(7)$ & $0$&$930(5)$  
&  $-0$&$258(19)$ & $0$&$198(6)$ \\
   $24$
&& $-1$&$474(6)$ & $-1$&$599(7)$ & $0$&$930(5)$  
&  $-0$&$257(19)$ & $0$&$201(8)$ \\
   $32$
&& $-1$&$473(7)$ & $-1$&$596(7)$ & $0$&$931(5)$  
&  $-0$&$270(20)$ & $0$&$191(9)$ \\
   {\it CL}
&& ${\it -1}$&${\it 471(8)}$ & ${\it -1}$&${\it 591(8)}$ 
&  ${\it 0}$&${\it 933(6)}$ & ${\it -0}$&${\it 271(23)}$ 
&  ${\it 0}$&${\it 197(9)}$      \\
\hline
\multicolumn{12}{c}{$z=9.0$}     \\
\hline
   $12$
&& $-1$&$56(1)$  & $-1$&$678(7)$ & $0$&$933(8)$  
&  $-0$&$166(31)$ & $0$&$149(4)$ \\
   $16$
&& $-1$&$530(8)$ & $-1$&$633(7)$ & $0$&$943(6)$  
&  $-0$&$182(27)$ & $0$&$172(5)$ \\
   $20$
&& $-1$&$522(6)$ & $-1$&$618(7)$ & $0$&$948(5)$  
&  $-0$&$190(25)$ & $0$&$176(6)$ \\
   $24$
&& $-1$&$520(6)$ & $-1$&$612(7)$ & $0$&$950(5)$  
&  $-0$&$189(25)$ & $0$&$180(8)$ \\
   $32$
&& $-1$&$517(7)$ & $-1$&$608(7)$ & $0$&$951(5)$  
&  $-0$&$200(26)$ & $0$&$170(9)$ \\
   {\it CL}
&& ${\it -1}$&${\it 515(11)}$ & ${\it -1}$&${\it 600(12)}$ 
&  ${\it 0}$&${\it 954(7)}$ & ${\it -0}$&${\it 202(30)}$ 
&  ${\it 0}$&${\it 177(9)}$      \\
\hline
\multicolumn{12}{c}{$z=13.5$}    \\
\hline
   $16$
&& $-1$&$639(8)$ & $-1$&$713(7)$ & $0$&$961(6)$  
&  $-0$&$106(47)$ & $0$&$139(5)$ \\
   $20$
&& $-1$&$595(7)$ & $-1$&$659(7)$ & $0$&$967(5)$  
&  $-0$&$123(39)$ & $0$&$149(5)$ \\
   $24$
&& $-1$&$583(7)$ & $-1$&$640(7)$ & $0$&$970(6)$  
&  $-0$&$126(37)$ & $0$&$157(8)$ \\
   $32$
&& $-1$&$578(7)$ & $-1$&$631(7)$ & $0$&$973(6)$  
&  $-0$&$132(39)$ & $0$&$142(9)$ \\
   {\it CL}
&& ${\it -1}$&${\it 571(17)}$ & ${\it -1}$&${\it 619(18)}$ 
&  ${\it 0}$&${\it 976(9)}$ & ${\it -0}$&${\it 137(66)}$ 
&  ${\it 0}$&${\it 145(14)}$     \\
\hline\hline
\end{tabular}
\caption{
(continued).
}\addtocounter{table}{-1}

}
%
\clearpage

\section{Perturbative conversion factors between QCD and HQET}
\label{App_Cx}
In this appendix we provide details on the numerical evaluation of the
perturbative approximations of the conversion functions 
$C_{\rm X}(M/\Lambda_{\rm \overline{MS}})$ (${\rm X=PS,V,PS/V,spin}$) that 
translate the matrix elements and energies obtained in the effective theory 
to those in {\em quenched} QCD at finite values of the heavy quark mass.
Also the case of the conversion factor 
$C_{\rm mass}(M/\Lambda_{\rm \overline{MS}})$, which relates the heavy 
quark's pole mass $m_Q$ to the renormalization group invariant quark mass 
$M$, will be addressed.

Let us begin with the definition of the conversion functions for matrix
elements of the heavy-light currents.
Parametrized with the $\overline{\rm MS}$ mass $\mbar_\ast$, implicitly 
defined through
\be \label{e_mstar}
 \mbar_\msbar(\mbar_\ast)=\mbar_\ast \,,
\ee
we write them for ${\rm X=\rm PS}$ and V as
\be
\widehat{C}_{\rm X}(\mbar_\ast)\equiv
\left[\,2b_0\bar g^2(\mbar_\ast)\,\right]^{\gamma_0^{\rm X}/(2b_0)}
\exp\left\{\int\limits_0^{\bar g(\mbar_\ast)} \rmd g 
\left[\,\frac{\gamma^{\rm X}(g)}{\beta(g)}-\frac{\gamma_0^{\rm X}}{b_0g}
\,\right]\right\}.
\label{Chat}
\ee
Here, $\beta(g)=-g^3 b_0-g^5b_1+\ldots$ is the four-loop anomalous 
dimension \cite{vanRitbergen:1997va} of the renormalized coupling 
$\bar g(\mu)$ in the $\rm \overline{MS}$ scheme with the leading- and 
next-to-leading-order coefficients $b_0={11}/{(4\pi)^2}$ and 
$b_1=102/(4\pi)^4$. 
In \eq{Chat} we have introduced the anomalous dimensions in the matching 
scheme
\be
\gamma^{\rm X}(g) = 
-g^2\left\{\,
\gamma_0^{\rm X}+\gamma_1^{\rm X} g^2 + \gamma_2^{\rm X} g^4 +
\ldots\,\right\}\,.
\label{ADgen}
\ee
At one-loop order they are the universal ones 
\cite{Shifman:1987sm,Politzer:1988wp},
\be 
 \gamma_0^{\rm PS} = \gamma_0^{\rm V}=-\frac{1}{4\pi^2}\,,
\ee
and at higher order they are related to the anomalous dimensions 
$\gamma^{\rm X,\,\msbar}(g)$ of the corresponding effective theory operator 
in the $\overline{\rm MS}$ scheme via
\bes 
 \gamma_1^{\rm X} &=& \gamma_1^{\rm X,\, \msbar} +2b_0c_1^{\rm X}\,, 
 \label{ADgen_1}\\
 \gamma_2^{\rm X} &=& \gamma_2^{\rm X,\, \msbar} 
                      +4b_0(c_2^{\rm X}+\gamma_0^{\rm X} k)
                      +2b_1c_1^{\rm X}-2b_0[c_1^{\rm X}]^2 \,.
 \label{ADgen_2} 
\ees
For ${\rm X=PS}$ and V, the $\overline{\rm MS}$ two-loop anomalous 
dimensions are known from \cite{Ji:1991pr,BroadhGrozin1,Gimenez:1992bf}
and the three-loop ones from \Ref{ChetGrozin}.\footnote{
Note that, since in HQET the anomalous dimension of the quark bilinears 
does not depend on their Dirac structure, 
$\gamma_n^{\rm PS,\, \msbar}=\gamma_n^{\rm V,\,\msbar}$ holds for all $n$.
}
The coefficients 
\bes
c_1^{\rm PS}   = -{2\over 3}\,{1\over 4\pi^2} \,,
& &   
c_2^{\rm PS}   = -4.2\,{1\over (4\pi^2)^2}  \,,\\
c_1^{\rm V}    = -{4\over 3}\,{1\over 4\pi^2} \,,
& & 
c_2^{\rm V}    = -11.5\,{1\over (4\pi^2)^2}
\ees
originate from the matching of the effective theory operators renormalized 
in the $\msbar$ scheme to the physical ones in QCD 
\cite{stat:eichhill1,stat:eichhill_za,BroadhGrozin2}, while the term 
proportional to 
\be
 k=-{1\over 3\pi^2}\,
 \label{m1loop}
\ee
is due to a reparametrization: the matching was originally done at the 
matching scale expressed in terms of the heavy quark's pole mass, $m_Q$. 
Using the perturbative expansion for the ratio 
$\mbar_\ast/m_Q$ \cite{polemass:1loop},
\be
 \frac{\mbar_\ast}{m_Q} = 1+ k\,\gbar^2(\mbar_\ast) + \ldots\,,
 \label{mratio}
\ee 
the pole mass can be replaced by $\mbar_\ast$.

Next, we turn to the chromomagnetic operator 
$\heavyb\vecsigma\cdot\vecB\,\heavy$.
If, for the moment, we follow the common practice that in the HQET 
expansion its matrix element is understood to multiply the inverse pole 
mass, $1/m_Q$, the associated conversion function $\Cmaghat$ would be 
given by \eq{Chat} with an expansion (\ref{ADgen}) for ${\rm X=mag}$ and
the universal one-loop coefficient \cite{stat:eichhill_1m,Falk:1991pz}
\be 
 \gamma_0^{\rm mag}=\frac{6}{(4\pi)^2}\,.
\ee
With the two-loop anomalous dimension $\gamma_1^{\rm mag,\,\msbar}$ in the 
effective theory given in \cite{HQET:sigmabI,HQET:sigmabII} one finds 
\bes
 \gamma_1^{\rm mag} = \gamma_1^{\rm mag,\, \msbar} +2b_0c_1^{\rm mag}\,,
 \quad
 c_1^{\rm mag}      = {13\over 6}\,{1\over 4\pi^2}\,,
 \label{ADmag_1} 
\ees
where $c_1^{\rm mag}$ was determined in \cite{stat:eichhill_1m}.
In view of \eq{e_XRGIspin}, however, we are rather interested in a function 
$\Cspin$, which multiplies $1/M$.
In other words, the conversion function $\Cspinhat$ must also include the 
factors $\mbar_\ast/m_Q$ and $M/\mbar_\ast$ in order to cancel the 
conventional factor $1/m_Q$ in the HQET expansion in favour of $1/M$.
Using the relation
\be
\frac{M}{\mbar_\ast}= 
\left[\,2b_0\bar g^2(\mbar_\ast)\,\right]^{-d_0/(2b_0)}
\exp\left\{-\int\limits_0^{\bar g(\mbar_\ast)} \rmd g 
\left[\,\frac{\tau(g)}{\beta(g)}-\frac{d_0}{b_0g}\,\right]\right\}\,,
\label{Movermbar}
\ee
where $\tau(g)=-g^2 d_0-g^4d_1+\ldots$ denotes the quark mass anomalous 
dimension in the $\msbar$ scheme in QCD
(known up to four loops \cite{Chetyrkin:1997dh,Vermaseren:1997fq}
with leading coefficient $d_0=8/(4\pi)^2$), and the ratio (\ref{mratio}), 
we then obtain \eqs{Chat} and (\ref{ADgen}) for ${\rm X=spin}$ and
\bes 
\gamma_0^{\rm spin} 
& = & 
\gamma_0^{\rm mag}-d_0=\frac{6}{(4\pi)^2}-d_0\,,\\
\gamma_1^{\rm spin} 
& = & 
\gamma_1^{\rm mag}-d_1+2b_0k=
\gamma_1^{\rm mag,\,\msbar}-d_1+2b_0(c_1^{\rm mag}+k)\,,
\ees
with $\gamma_1^{\rm mag}$ from \eq{ADmag_1}.

For the special case of the ratio of pseudoscalar and vector current matrix
elements, ${\rm X=PS/V}$, all but the contributions from the matching 
cancel and one gets 
\be
\widehat{C}_{\rm PS/V}(\mbar_\ast)\equiv
\exp\left\{\int\limits_0^{\bar g(\mbar_\ast)} \rmd g\,\,
\frac{\gamma^{\rm PS}(g)-\gamma^{\rm V}(g)}{\beta(g)}\right\}\,.
\ee

To parametrize the mass dependence of energy observables (such as $\gamav$ 
in \eq{e_effenergy_exp} of \sect{s_ener}), we also define 
\be
\widehat{C}_{\rm mass}(\mbar_\ast)\equiv
{m_Q\over \mbar_\ast}\,{\mbar_\ast\over M}\,,
\ee
where in this case the highest available perturbative precision is 
achieved by taking the four-loop 
$\tau$-function \cite{Chetyrkin:1997dh,Vermaseren:1997fq} 
in $\mbar_\ast/M$ together with the three-loop expression for 
$m_Q/\mbar_\ast$ \cite{polemass:3loop}.

Finally, changing the argument of the various $\widehat{C}_{X}$ to the
renormalization group invariant ratio $M/\Lambda_{\rm \overline{MS}}$
via (\ref{Movermbar}), we straightforwardly arrive at the conversion 
functions
\be
C_{\rm X}\left(M/\Lambda_{\msbar}\right) = 
\widehat{C}_{\rm X}\left(\mbar_\ast\right)
\quad{\rm with}\quad{\rm X=PS,V,PS/V,mass,spin}\,.
\ee
We evaluate all occurring integrals in the above expressions exactly 
(sometimes numerically), truncating anomalous dimensions and 
the $\beta$-function as specified.
Also \eq{e_mstar} is solved numerically. 
For practical purposes, such as repeated use in the fits of the heavy 
quark mass dependence of our QCD observables, a parametrization of all 
conversion functions in terms of the variable 
\be
x\equiv
\frac{1}{\ln\left(\M/\Lambda_{\msbar}\right)}
\ee
was determined from a numerical evaluation.
The functions decompose into a prefactor encoding the leading asymptotics 
as $x\rightarrow0$, multiplied by a polynomial in $x$, which guarantees 
at least 0.2\% precision for $x\leq 0.6$:
\bea
\Cps(x)
& = &
\left\{
\begin{array}{ll}
x^{\gamma_0^{\rm PS}/(2b_0)}\left\{\,1-0.065\,x+0.048\,x^2\,\right\}
& \quad \mbox{2-loop $\gamma^{\rm PS}$}\\
\\
x^{\gamma_0^{\rm PS}/(2b_0)}
\left\{\,1-0.068\,x-0.087\,x^2+0.079\,x^3\,\right\}
& \quad \mbox{3-loop $\gamma^{\rm PS}$}
\end{array}
\right. \,,
\label{e_cpsfit}\\
\nonumber\\ \nonumber\\[-1ex]
\Cv(x)
& = &
\left\{
\begin{array}{ll}
x^{\gamma_0^{\rm V}/(2b_0)}\left\{\,1-0.180\,x+0.099\,x^2\,\right\}
& \quad \mbox{2-loop $\gamma^{\rm V}$}\\
\\
x^{\gamma_0^{\rm V}/(2b_0)}
\left\{\,1-0.196\,x-0.222\,x^2+0.193\,x^3\,\right\}
& \quad \mbox{3-loop $\gamma^{\rm V}$}
\end{array}
\right. \,,
\label{e_cvfit}\\
\nonumber\\ \nonumber\\[-1ex]
\Cpsv(x)
& = &
\left\{
\begin{array}{ll}
1+0.117\,x-0.043\,x^2
& \quad \mbox{2-loop $\gamma^{\rm PS,V}$}\\
\\
1+0.124\,x+0.187\,x^2-0.102\,x^3
& \quad \mbox{3-loop $\gamma^{\rm PS,V}$}
\end{array}
\right. \,,
\label{e_cpsvfit}\\
\nonumber\\ \nonumber\\[-1ex]
\Cmass(x)
& = &
\left\{
\begin{array}{ll}
x^{d_0/(2b_0)}\left\{\,1+0.247\,x+0.236\,x^2\,\right\}
& \quad \mbox{2-loop $\frac{m_Q}{\mbar_\ast}$}\\
\\
x^{d_0/(2b_0)}\left\{\,1+0.179\,x+0.694\,x^2+0.065\,x^3\,\right\}
& \quad \mbox{3-loop $\frac{m_Q}{\mbar_\ast}$}
\end{array}
\right. \,,
\label{e_cmassfit}\\
\nonumber\\ \nonumber\\[-1ex]
\Cspin(x)
& = &
\left\{
\begin{array}{ll}
x^{\gamma_0^{\rm spin}/(2b_0)}\left\{\,1+0.066\,x\,\right\}
& \quad \mbox{1-loop $\gamma^{\rm spin}$}\\
\\
x^{\gamma_0^{\rm spin}/(2b_0)}\left\{\,1+0.087\,x-0.021\,x^2\,\right\}
& \quad \mbox{2-loop $\gamma^{\rm spin}$}
\end{array}
\right. \,.
\label{e_cspinfit}
\eea
\begin{minipage}{\textwidth}
Apart from the function $\Cmass$, the pole mass does not appear in any of
the above perturbative expressions; they relate observables in the 
effective theory to those in QCD and are parametrized by the RGI mass $M$,
which is unambiguously defined in terms of the (short-distance) running 
quark mass (see \eq{e_M}). 
Thus, their perturbative expansion is expected to be a regular 
short-distance expansion. 
In particular, it is not expected to suffer from the bad behaviour of the 
series (\ref{mratio}).
\end{minipage}
\end{appendix}
\bibliography{lattice_ALPHA}

\providecommand{\href}[2]{#2}\begingroup\raggedright\begin{thebibliography}{10}

\bibitem{stat:eichhill1}
E.~Eichten and B.~Hill, {\it An effective field theory for the calculation of
  matrix elements involving heavy quarks},  {\em Phys. Lett.} {\bf B234} (1990)
  511.

\bibitem{hqet:polwis}
H.~D. Politzer and M.~B. Wise, {\it Effective field theory approach to
  processes involving both light and heavy fields},  {\em Phys. Lett.} {\bf
  B208} (1988) 504.

\bibitem{hqet:georgi}
H.~Georgi, {\it An effective field theory for heavy quarks at low energies},
  {\em Phys. Lett.} {\bf B240} (1990) 447.

\bibitem{HQET:neubert}
M.~Neubert, {\it Heavy quark symmetry},  {\em Phys. Rept.} {\bf 245} (1994) 259
  [\href{http://arXiv.org/abs/hep-ph/9306320}{{\tt hep-ph/9306320}}].

\bibitem{reviews:HQETMannel}
T.~Mannel, {\it Effective theory for heavy quarks},  {\em Lectures at 35th
  Internationale {Universit\"atswochen} {f\"ur} Kern- und Teilchenphysik,
  Schladming, Austria, 2 -- 9 March 1996}
  [\href{http://arXiv.org/abs/hep-ph/9606299}{{\tt hep-ph/9606299}}].

\bibitem{HQET:renormI}
B.~Grinstein, {\it The static quark effective theory},  {\em Nucl. Phys.} {\bf
  B339} (1990) 253.

\bibitem{HQET:renormII}
W.~Kilian and T.~Mannel, {\it On the renormalization of heavy quark effective
  field theory},  {\em Phys. Rev.} {\bf D49} (1994) 1534
  [\href{http://arXiv.org/abs/hep-ph/9307307}{{\tt hep-ph/9307307}}].

\bibitem{reviews:HQETpert}
A.~G. Grozin, {\it Lectures on perturbative {HQET}.~{I}},
  \href{http://arXiv.org/abs/hep-ph/0008300}{{\tt hep-ph/0008300}}.

\bibitem{ChetGrozin}
K.~G. Chetyrkin and A.~G. Grozin, {\it Three-loop anomalous dimension of the
  heavy-light quark current in {HQET}},  {\em Nucl. Phys.} {\bf B666} (2003)
  289 [\href{http://arXiv.org/abs/hep-ph/0303113}{{\tt hep-ph/0303113}}].

\bibitem{PDGlast}
{\bf Particle Data Group} Collaboration, K.~Hagiwara {\em et.~al.}, {\it Review
  of particle physics},  {\em Phys. Rev.} {\bf D66} (2002) 010001
  [\href{http://arXiv.org/abs/http://pdg.lbl.gov}{{\tt http://pdg.lbl.gov}}].

\bibitem{reviews:Stone03}
S.~Stone, {\it Experimental results in heavy flavor physics},  {\em Plenary
  talk at International Europhysics Conference on High-Energy Physics (HEP
  2003), Aachen, Germany, 17 -- 23 July 2003, Eur. Phys. J.} {\bf C33} (2004)
  S129 [\href{http://arXiv.org/abs/hep-ph/0310153}{{\tt hep-ph/0310153}}].

\bibitem{CKM:CERN}
M.~Battaglia {\em et.~al.}, {\it The {CKM} matrix and the unitarity triangle},
  {\em Proceedings of the {CKM} Unitarity Triangle Workshop, Geneva,
  Switzerland, 13 -- 16 February 2002}
  [\href{http://arXiv.org/abs/hep-ph/0304132}{{\tt hep-ph/0304132}}].

\bibitem{Babar:hqetpar}
{\bf BABAR} Collaboration, B.~Aubert {\em et.~al.}, {\it Determination of the
  branching fraction for ${B}\to {X}_{c} l \nu$ decays and of $|{V}_{cb}|$ from
  hadronic mass and lepton energy moments},  {\em Phys. Rev. Lett.} {\bf 93}
  (2004) 011803 [\href{http://arXiv.org/abs/hep-ex/0404017}{{\tt
  hep-ex/0404017}}].

\bibitem{zastat:pap2}
{\bf ALPHA} Collaboration, M.~Kurth and R.~Sommer, {\it Heavy quark effective
  theory at one-loop order: An explicit example},  {\em Nucl. Phys.} {\bf B623}
  (2002) 271 [\href{http://arXiv.org/abs/hep-lat/0108018}{{\tt
  hep-lat/0108018}}].

\bibitem{Bernard:2001fz}
C.~Bernard {\em et.~al.}, {\it Lattice results for the decay constant of
  heavy-light vector mesons},  {\em Phys. Rev.} {\bf D65} (2002) 014510
  [\href{http://arXiv.org/abs/hep-lat/0109015}{{\tt hep-lat/0109015}}].

\bibitem{mcbar:RS02}
{\bf ALPHA} Collaboration, J.~Rolf and S.~Sint, {\it A precise determination of
  the charm quark's mass in quenched {QCD}},  {\em J. High Energy Phys.} {\bf
  12} (2002) 007 [\href{http://arXiv.org/abs/hep-ph/0209255}{{\tt
  hep-ph/0209255}}].

\bibitem{fds:JR03}
{\bf ALPHA} Collaboration, A.~{J\"uttner} and J.~Rolf, {\it A precise
  determination of the decay constant of the ${D}_s$ meson in quenched {QCD}},
  {\em Phys. Lett.} {\bf B560} (2003) 59
  [\href{http://arXiv.org/abs/hep-lat/0302016}{{\tt hep-lat/0302016}}].

\bibitem{lat03:Juri}
{\bf ALPHA} Collaboration, J.~Rolf, M.~{Della Morte}, S.~{D\"urr}, J.~Heitger,
  A.~{J\"uttner}, H.~Molke, A.~Shindler and R.~Sommer, {\it Towards a precision
  computation of ${F}_{B_s}$ in quenched {QCD}},  {\em Nucl. Phys. Proc.
  Suppl.} {\bf 129} (2004) 322
  [\href{http://arXiv.org/abs/hep-lat/0309072}{{\tt hep-lat/0309072}}].

\bibitem{fb_wupp}
C.~Alexandrou {\em et.~al.}, {\it The leptonic decay constants of $\bar{{Q}}q$
  mesons and the lattice resolution},  {\em Z. Phys.} {\bf C62} (1994) 659
  [\href{http://arXiv.org/abs/hep-lat/9312051}{{\tt hep-lat/9312051}}].

\bibitem{reviews:beauty}
R.~Sommer, {\it Beauty physics in lattice gauge theory},  {\em Phys. Rept.}
  {\bf 275} (1996) 1 [\href{http://arXiv.org/abs/hep-lat/9401037}{{\tt
  hep-lat/9401037}}].

\bibitem{reviews:hartmut97}
H.~Wittig, {\it Leptonic decays of heavy quarks on the lattice},  {\em Int. J.
  Mod. Phys.} {\bf A12} (1997) 4477
  [\href{http://arXiv.org/abs/hep-lat/9705034}{{\tt hep-lat/9705034}}].

\bibitem{El-Khadra:1998hq}
A.~X. El-Khadra, A.~S. Kronfeld, P.~B. Mackenzie, S.~M. Ryan and J.~N. Simone,
  {\it {B} and {D} meson decay constants in lattice {QCD}},  {\em Phys. Rev.}
  {\bf D58} (1998) 014506 [\href{http://arXiv.org/abs/hep-ph/9711426}{{\tt
  hep-ph/9711426}}].

\bibitem{Aoki:1998ji}
{\bf JLQCD} Collaboration, S.~Aoki {\em et.~al.}, {\it Heavy meson decay
  constants from quenched lattice {QCD}},  {\em Phys. Rev. Lett.} {\bf 80}
  (1998) 5711.

\bibitem{Bernard:1998xi}
C.~W. Bernard {\em et.~al.}, {\it Lattice determination of heavy-light decay
  constants},  {\em Phys. Rev. Lett.} {\bf 81} (1998) 4812
  [\href{http://arXiv.org/abs/hep-ph/9806412}{{\tt hep-ph/9806412}}].

\bibitem{heavylight:Bec98}
D.~Becirevic {\em et.~al.}, {\it Non-perturbatively improved heavy-light
  mesons: {M}asses and decay constants},  {\em Phys. Rev.} {\bf D60} (1999)
  074501 [\href{http://arXiv.org/abs/hep-lat/9811003}{{\tt hep-lat/9811003}}].

\bibitem{Bowler:2000xw}
{\bf UKQCD} Collaboration, K.~C. Bowler {\em et.~al.}, {\it Decay constants of
  {B} and {D} mesons from non-perturbatively improved lattice {QCD}},  {\em
  Nucl. Phys.} {\bf B619} (2001) 507
  [\href{http://arXiv.org/abs/hep-lat/0007020}{{\tt hep-lat/0007020}}].

\bibitem{AliKhan:2000eg}
{\bf CP-PACS} Collaboration, A.~Ali~Khan {\em et.~al.}, {\it Decay constants of
  {B} and {D} mesons from improved relativistic lattice {QCD} with two flavours
  of sea quarks},  {\em Phys. Rev.} {\bf D64} (2001) 034505
  [\href{http://arXiv.org/abs/hep-lat/0010009}{{\tt hep-lat/0010009}}].

\bibitem{Lellouch:2000tw}
{\bf UKQCD} Collaboration, L.~Lellouch and C.~J.~D. Lin, {\it Standard model
  matrix elements for neutral {B} meson mixing and associated decay constants},
   {\em Phys. Rev.} {\bf D64} (2001) 094501
  [\href{http://arXiv.org/abs/hep-ph/0011086}{{\tt hep-ph/0011086}}].

\bibitem{lat01:ryan}
S.~M. Ryan, {\it Heavy quark physics from lattice {QCD}},  {\em Nucl. Phys.
  Proc. Suppl.} {\bf 106} (2002) 86
  [\href{http://arXiv.org/abs/hep-lat/0111010}{{\tt hep-lat/0111010}}].

\bibitem{mb:roma2}
G.~M. de~Divitiis, M.~Guagnelli, R.~Petronzio, N.~Tantalo and F.~Palombi, {\it
  Heavy quark masses in the continuum limit of lattice {QCD}},  {\em Nucl.
  Phys.} {\bf B675} (2003) 309
  [\href{http://arXiv.org/abs/hep-lat/0305018}{{\tt hep-lat/0305018}}].

\bibitem{fb:roma2c}
G.~M. de~Divitiis, M.~Guagnelli, F.~Palombi, R.~Petronzio and N.~Tantalo, {\it
  Heavy-light decay constants in the continuum limit of lattice {QCD}},  {\em
  Nucl. Phys.} {\bf B672} (2003) 372
  [\href{http://arXiv.org/abs/hep-lat/0307005}{{\tt hep-lat/0307005}}].

\bibitem{SF:LNWW}
M.~{L\"uscher}, R.~Narayanan, P.~Weisz and U.~Wolff, {\it The {Schr\"odinger}
  functional: A renormalizable probe for non-abelian gauge theories},  {\em
  Nucl. Phys.} {\bf B384} (1992) 168
  [\href{http://arXiv.org/abs/hep-lat/9207009}{{\tt hep-lat/9207009}}].

\bibitem{SF:stefan1}
S.~Sint, {\it On the {Schr\"odinger} functional in {QCD}},  {\em Nucl. Phys.}
  {\bf B421} (1994) 135 [\href{http://arXiv.org/abs/hep-lat/9312079}{{\tt
  hep-lat/9312079}}].

\bibitem{lat01:mbstat}
{\bf ALPHA} Collaboration, J.~Heitger and R.~Sommer, {\it A strategy to compute
  the b-quark mass with non-perturbative accuracy},  {\em Nucl. Phys. Proc.
  Suppl.} {\bf 106} (2002) 358
  [\href{http://arXiv.org/abs/hep-lat/0110016}{{\tt hep-lat/0110016}}].

\bibitem{HQET:pap2}
{\bf ALPHA} Collaboration, J.~Heitger and J.~Wennekers, {\it Effective
  heavy-light meson energies in small-volume quenched {QCD}},  {\em J. High
  Energy Phys.} {\bf 02} (2004) 064
  [\href{http://arXiv.org/abs/hep-lat/0312016}{{\tt hep-lat/0312016}}].

\bibitem{HQET:pap1}
{\bf ALPHA} Collaboration, J.~Heitger and R.~Sommer, {\it Non-perturbative
  heavy quark effective theory},  {\em J. High Energy Phys.} {\bf 02} (2004)
  022 [\href{http://arXiv.org/abs/hep-lat/0310035}{{\tt hep-lat/0310035}}].

\bibitem{zastat:pap1}
{\bf ALPHA} Collaboration, M.~Kurth and R.~Sommer, {\it Renormalization and
  {O}($a$) improvement of the static axial current},  {\em Nucl. Phys.} {\bf
  B597} (2001) 488 [\href{http://arXiv.org/abs/hep-lat/0007002}{{\tt
  hep-lat/0007002}}].

\bibitem{SF:stefan2}
S.~Sint, {\it One-loop renormalization of the {QCD} {Schr\"odinger}
  functional},  {\em Nucl. Phys.} {\bf B451} (1995) 416
  [\href{http://arXiv.org/abs/hep-lat/9504005}{{\tt hep-lat/9504005}}].

\bibitem{zastat:pap3}
{\bf ALPHA} Collaboration, J.~Heitger, M.~Kurth and R.~Sommer, {\it
  Non-perturbative renormalization of the static axial current in quenched
  {QCD}},  {\em Nucl. Phys.} {\bf B669} (2003) 173
  [\href{http://arXiv.org/abs/hep-lat/0302019}{{\tt hep-lat/0302019}}].

\bibitem{impr:pap3}
{\bf ALPHA} Collaboration, M.~{L\"uscher}, S.~Sint, R.~Sommer, P.~Weisz and
  U.~Wolff, {\it Non-perturbative {O}($a$) improvement of lattice {QCD}},  {\em
  Nucl. Phys.} {\bf B491} (1997) 323
  [\href{http://arXiv.org/abs/hep-lat/9609035}{{\tt hep-lat/9609035}}].

\bibitem{msbar:pap1}
{\bf ALPHA} Collaboration, S.~Capitani, M.~{L\"uscher}, R.~Sommer and
  H.~Wittig, {\it Non-perturbative quark mass renormalization in quenched
  lattice {QCD}},  {\em Nucl. Phys} {\bf B544} (1999) 669
  [\href{http://arXiv.org/abs/hep-lat/9810063}{{\tt hep-lat/9810063}}].

\bibitem{impr:babp}
{\bf ALPHA} Collaboration, M.~Guagnelli, R.~Petronzio, J.~Rolf, S.~Sint,
  R.~Sommer and U.~Wolff, {\it Non-perturbative results for the coefficients
  $b_m$ and $b_{A}-b_{P}$ in {O}($a$) improved lattice {QCD}},  {\em Nucl.
  Phys.} {\bf B595} (2001) 44 [\href{http://arXiv.org/abs/hep-lat/0009021}{{\tt
  hep-lat/0009021}}].

\bibitem{vanRitbergen:1997va}
T.~van Ritbergen, J.~A.~M. Vermaseren and S.~A. Larin, {\it The four-loop
  $\beta$-function in {Q}uantum {C}hromodynamics},  {\em Phys. Lett.} {\bf
  B400} (1997) 379 [\href{http://arXiv.org/abs/hep-ph/9701390}{{\tt
  hep-ph/9701390}}].

\bibitem{HQET:massform2}
A.~F. Falk and M.~Neubert, {\it Second-order power corrections in the heavy
  quark effective theory. 1.~{F}ormalism and meson form-factors},  {\em Phys.
  Rev.} {\bf D47} (1993) 2965 [\href{http://arXiv.org/abs/hep-ph/9209268}{{\tt
  hep-ph/9209268}}].

\bibitem{Chetyrkin:1997dh}
K.~G. Chetyrkin, {\it Quark mass anomalous dimension to {O}($\alpha_s^4$)},
  {\em Phys. Lett.} {\bf B404} (1997) 161
  [\href{http://arXiv.org/abs/hep-ph/9703278}{{\tt hep-ph/9703278}}].

\bibitem{Vermaseren:1997fq}
J.~A.~M. Vermaseren, S.~A. Larin and T.~van Ritbergen, {\it The four-loop quark
  mass anomalous dimension and the invariant quark mass},  {\em Phys. Lett.}
  {\bf B405} (1997) 327 [\href{http://arXiv.org/abs/hep-ph/9703284}{{\tt
  hep-ph/9703284}}].

\bibitem{polemass:3loop}
N.~Gray, D.~J. Broadhurst, W.~Grafe and K.~Schilcher, {\it Three-loop relation
  of quark (modified) $\overline{MS}$ and pole masses},  {\em Z. Phys.} {\bf
  C48} (1990) 673.

\bibitem{HQET:reparaI}
M.~E. Luke and A.~V. Manohar, {\it Reparametrization invariance constraints on
  heavy particle effective field theories},  {\em Phys. Lett.} {\bf B286}
  (1992) 348 [\href{http://arXiv.org/abs/hep-ph/9205228}{{\tt
  hep-ph/9205228}}].

\bibitem{HQET:reparaII}
W.~Kilian and T.~Ohl, {\it Renormalization of heavy quark effective field
  theory: Quantum action principles and equations of motion},  {\em Phys. Rev.}
  {\bf D50} (1994) 4649 [\href{http://arXiv.org/abs/hep-ph/9404305}{{\tt
  hep-ph/9404305}}].

\bibitem{HQET:reparaIII}
R.~Sundrum, {\it Reparameterization invariance to all orders in heavy quark
  effective theory},  {\em Phys. Rev.} {\bf D57} (1998) 331
  [\href{http://arXiv.org/abs/hep-ph/9704256}{{\tt hep-ph/9704256}}].

\bibitem{HQET:sigmabI}
G.~Amoros, M.~Beneke and M.~Neubert, {\it Two-loop anomalous dimension of the
  chromomagnetic moment of a heavy quark},  {\em Phys. Lett.} {\bf B401} (1997)
  81 [\href{http://arXiv.org/abs/hep-ph/9701375}{{\tt hep-ph/9701375}}].

\bibitem{HQET:sigmabII}
A.~Czarnecki and A.~G. Grozin, {\it {HQET} chromomagnetic interaction at two
  loops},  {\em Phys. Lett.} {\bf B405} (1997) 142
  [\href{http://arXiv.org/abs/hep-ph/9701415}{{\tt hep-ph/9701415}}].

\bibitem{alpha:SU3}
M.~{L\"uscher}, R.~Sommer, P.~Weisz and U.~Wolff, {\it A precise determination
  of the running coupling in the {SU(3)} {Y}ang-{M}ills theory},  {\em Nucl.
  Phys.} {\bf B413} (1994) 481
  [\href{http://arXiv.org/abs/hep-lat/9309005}{{\tt hep-lat/9309005}}].

\bibitem{stat:symm1}
N.~Isgur and M.~B. Wise, {\it Weak decays of heavy mesons in the static quark
  approximation},  {\em Phys. Lett.} {\bf B232} (1989) 113.

\bibitem{stat:symm2}
N.~Isgur and M.~B. Wise, {\it Weak transition form-factors between heavy
  mesons},  {\em Phys. Lett.} {\bf B237} (1990) 527.

\bibitem{fbstat:pap1}
{\bf ALPHA} Collaboration, M.~{Della Morte}, S.~{D\"urr}, J.~Heitger, H.~Molke,
  J.~Rolf, A.~Shindler and R.~Sommer, {\it Lattice {HQET} with exponentially
  improved statistical precision},  {\em Phys. Lett.} {\bf B581} (2004) 93
  [\href{http://arXiv.org/abs/hep-lat/0307021}{{\tt hep-lat/0307021}}].

\bibitem{HYP:HK01}
A.~Hasenfratz and F.~Knechtli, {\it Flavor symmetry and the static potential
  with hypercubic blocking},  {\em Phys. Rev.} {\bf D64} (2001) 034504
  [\href{http://arXiv.org/abs/hep-lat/0103029}{{\tt hep-lat/0103029}}].

\bibitem{BroadhGrozin2}
D.~J. Broadhurst and A.~G. Grozin, {\it Matching {QCD} and {HQET} heavy-light
  currents at two loops and beyond},  {\em Phys. Rev.} {\bf D52} (1995) 4082
  [\href{http://arXiv.org/abs/hep-ph/9410240}{{\tt hep-ph/9410240}}].

\bibitem{code:MILC}
 \href{http://arXiv.org/abs/http://www.physics.indiana.edu/\~{}sg/milc.html}{{%
\tt http://www.physics.indiana.edu/\~{}sg/milc.html}}.

\bibitem{Shifman:1987sm}
M.~A. Shifman and M.~B. Voloshin, {\it On annihilation of mesons built from
  heavy and light quark and $\bar{B}_0 \leftrightarrow {B}_0$ oscillations},
  {\em Sov. J. Nucl. Phys.} {\bf 45} (1987) 292.

\bibitem{Politzer:1988wp}
H.~D. Politzer and M.~B. Wise, {\it Leading logarithms of heavy quark masses in
  processes with light and heavy quarks},  {\em Phys. Lett.} {\bf B206} (1988)
  681.

\bibitem{Ji:1991pr}
X.~Ji and M.~J. Musolf, {\it Subleading logarithmic mass dependence in heavy
  meson form-factors},  {\em Phys. Lett.} {\bf B257} (1991) 409.

\bibitem{BroadhGrozin1}
D.~J. Broadhurst and A.~G. Grozin, {\it Two-loop renormalization of the
  effective field theory of a static quark},  {\em Phys. Lett.} {\bf B267}
  (1991) 105 [\href{http://arXiv.org/abs/hep-ph/9908362}{{\tt
  hep-ph/9908362}}].

\bibitem{Gimenez:1992bf}
V.~Gimenez, {\it Two-loop calculation of the anomalous dimension of the axial
  current with static heavy quarks},  {\em Nucl. Phys.} {\bf B375} (1992) 582.

\bibitem{stat:eichhill_za}
E.~Eichten and B.~Hill, {\it Renormalization of heavy-light bilinears and
  $f_{B}$ for {W}ilson fermions},  {\em Phys. Lett.} {\bf B240} (1990) 193.

\bibitem{polemass:1loop}
R.~Tarrach, {\it The pole mass in perturbative {QCD}},  {\em Nucl. Phys.} {\bf
  B183} (1981) 384.

\bibitem{stat:eichhill_1m}
E.~Eichten and B.~Hill, {\it Static effective field theory: $1/m$ corrections},
   {\em Phys. Lett.} {\bf B243} (1990) 427.

\bibitem{Falk:1991pz}
A.~F. Falk, B.~Grinstein and M.~E. Luke, {\it Leading mass corrections to the
  heavy quark effective theory},  {\em Nucl. Phys.} {\bf B357} (1991) 185.

\end{thebibliography}\endgroup
\bibliographystyle{JHEP-2}
\end{document}